\documentclass{aa}

\usepackage{graphicx}
\usepackage{txfonts}
\usepackage{amssymb}
\usepackage{amsmath}
\usepackage{graphicx}
\usepackage[colorlinks=true, allcolors=blue]{hyperref}
\usepackage{booktabs}
\usepackage{amsfonts}     
\usepackage[T1]{fontenc} 
\usepackage[utf8]{inputenc} 
\usepackage{multicol}
\usepackage{xcolor}
\usepackage[switch]{lineno}
\usepackage{gensymb}
\usepackage{soul}
\usepackage{rotating}
\usepackage{comment}
\usepackage{multirow}
\usepackage{longtable}
\usepackage{lscape}

\newcommand{\eqb}{\begin{eqnarray}}
\newcommand{\eqe}{\end{eqnarray}}

\begin{document}

\title{Association of the IceCube neutrinos with blazars in the CGRaBS sample\thanks{This paper is accompanied by three tables \texttt{Neut.dat}, \texttt{Blz.dat}, and \texttt{Assoc.dat}, which are only available in electronic form at the CDS via anonymous ftp to cdsarc.u-strasbg.fr (130.79.128.5) or via http://cdsweb.u-strasbg.fr/cgi-bin/qcat?J/A+A/.}}

\subtitle{}

\author{
Pouya M. Kouch \inst{\ref{UTU},\ref{FINCA},\ref{MRO}} \thanks{\href{mailto:pouya.kouch@utu.fi}{pouya.kouch@utu.fi}} \orcid{0000-0002-9328-2750}
Elina Lindfors \inst{\ref{UTU},\ref{FINCA}} \orcid{0000-0002-9155-6199} 
Talvikki Hovatta \inst{\ref{FINCA},\ref{MRO}} \orcid{0000-0002-2024-8199} 
Ioannis Liodakis \inst{\ref{FINCA},\ref{NASA_Alabama}} \orcid{0000-0001-9200-4006} 
Karri I.I. Koljonen \inst{\ref{NTNU}} \orcid{0000-0002-9677-1533} 
Kari Nilsson \inst{\ref{FINCA}} \orcid{0000-0002-1445-8683} 
Sebastian Kiehlmann \inst{\ref{IA_FORTH_Crete},\ref{UniCrete}} \orcid{0000-0001-6314-9177} 
Walter Max-Moerbeck \inst{\ref{UniChile}} \orcid{0000-0002-5491-5244} 
Anthony C.S. Readhead \inst{\ref{OVRO}} \orcid{0000-0001-9152-961X} 
Rodrigo A. Reeves \inst{\ref{UniConcepcion}} \orcid{0000-0001-5704-271X} 
Timothy J. Pearson \inst{\ref{OVRO}} \orcid{0000-0001-5213-6231} 
Jenni Jormanainen \inst{\ref{UTU},\ref{FINCA}} \orcid{0000-0003-4519-7751} 
Vandad Fallah Ramazani \inst{\ref{Bochum}} \orcid{0000-0001-8991-7744} 
Matthew J. Graham \inst{\ref{CalTech}} \orcid{0000-0002-3168-0139} 
}

\institute{
Department of Physics and Astronomy, University of Turku, FI-20014, Finland \label{UTU}
\and
Finnish Centre for Astronomy with ESO (FINCA), Quantum, Vesilinnantie 5, FI-20014 University of Turku, Finland \label{FINCA}
\and
Aalto University Mets\"ahovi Radio Observatory, Mets\"ahovintie 114, FI-02540 Kylm\"al\"a, Finland \label{MRO}
\and
NASA Marshall Space Flight Center, Huntsville, AL 35812, USA \label{NASA_Alabama}
\and
Institutt for Fysikk, Norwegian University of Science and Technology, H{\o}gskloreringen 5, Trondheim, 7491, Norway \label{NTNU}
\and
Institute of Astrophysics, Foundation for Research and Technology-Hellas, GR-71110 Heraklion, Greece \label{IA_FORTH_Crete}
\and
Department of Physics, Univ. of Crete, GR-70013 Heraklion, Greece \label{UniCrete}
\and
Departamento de Astronom\'{i}a, Universidad de Chile, Camino El Observatorio 1515, Las Condes, Santiago, Chile \label{UniChile}
\and
Owens Valley Radio Observatory, California Institute of Technology, Pasadena, CA 91125, USA \label{OVRO}
\and
CePIA, Astronomy Department, Universidad de Concepci\'{o}n, Casilla 160-C, Concepci\'{o}n, Chile \label{UniConcepcion}
\and
Ruhr-Universit\"at Bochum, Fakult\"at f\"ur Physik und Astronomie, Astronomisches Institut (AIRUB), 44801 Bochum, Germany \label{Bochum}
\and
Division of Physics, Mathematics and Astronomy, California Institute of Technology, Pasadena, CA91125, USA \label{CalTech}
}

\date{Received August 01, 2023; accepted July 09, 2024}

 
\abstract{The origin of high-energy (HE) astrophysical neutrinos has remained an elusive hot topic in the field of HE astrophysics for the past decade. Apart from a handful of individual associations, the vast majority of HE neutrinos arise from unknown sources. While there are theoretically-motivated candidate populations, such as blazars -- a subclass of AGN with jets pointed towards our line-of-sight -- they have not yet been convincingly linked to HE neutrino production. Here, we perform a spatio-temporal association analysis between a sample of blazars (from CGRaBS catalog) in the radio and optical bands and the most up-to-date IceCube HE neutrino catalog. We find that if the IceCube error regions are enlarged by 1$^\circ$ in quadrature, to account for unknown systematic errors at maximal level, a spatio-temporal correlation between the multiwavelength light curves of the CGRaBS blazars and the IceCube HE neutrinos is hinted at least at a 2.17$\sigma$ significance level. On the other hand, when the IceCube error regions are taken as their published values, we do not find any significant correlations. A discrepancy in the blazar-neutrino correlation strengths, when using such minimal and enlarged error region scenarios, was also obtained in a recent study by the IceCube collaboration. In our study, this difference arises because several flaring blazars -- coinciding with a neutrino arrival time -- happen to narrowly miss the published 90\%-likelihood error region of the nearest neutrino event. For all of the associations driving our most significant correlations, the flaring blazar is much less than 1$^\circ$ away from the published error regions. Therefore, our results indicate that the question of the blazar-neutrino connection is highly sensitive to the reconstruction of the neutrino error regions, whose reliability is expected to improve with the next generation of neutrino observatories.}


\titlerunning{CGRaBS blazar-neutrino association in the radio and optical bands}
\authorrunning{Kouch et al.}

\maketitle
%

\section{Introduction} \label{sec:intro}
Up to now the IceCube neutrino observatory has detected hundreds of high-energy (HE, $\ge$100 TeV) neutrinos in the TeV-PeV energy range \citep{abbasi+2023_cat1}.
Despite the astrophysical origin of many HE neutrinos, it is still a mystery what sources and mechanisms are involved in their emission. Due to a strong atmospheric background and limited directional accuracy, the task of associating them with astrophysical objects is challenging (for a recent review see e.g. \citealt{troitsky2021}).

One of the main neutrino candidate sources are a subclass of Active Galactic Nuclei (AGN) called blazars.
Blazars possess highly collimated relativistic jets that happen to point almost directly towards Earth.
Their energy output is dominated by relativistically-boosted non-thermal emission in all bands, from radio to very high-energy (VHE) $\gamma$-rays.
For recent reviews on blazars see e.g. \cite{blandford+2019},  \cite{Bottcher19}, and \cite{hovatta&lindfors2019}.

In blazar jets, the expected particle energies and the external photon field density may lead to the creation of HE neutrinos via hadronic interactions (e.g. \citealt{mannheim89, mannheim+1992, gaisser+1995, mucke01, becker2008}) or possibly leptonic ones \citep{hooper&kathryn2023_leptonic_neut_production}.
Although blazar jets are a suspect birthplace of neutrinos, not all HE neutrinos come from blazars; for example, the nearby Seyfert II AGN, NGC~1068, was recently found to be the first significant persistent HE neutrino emitting source \citep{icecube_collab2022_ngc1068}.

On a source-by-source basis, the most famous spatio-temporal blazar-neutrino association is with the blazar TXS~0506+056 which was in a flaring state at the time of the neutrino arrival \citep{icecube_collab2018}.
Several population-based studies have attempted to investigate the blazar-neutrino connection with their results so far generally remaining inconclusive or, at best, suggestive (e.g. \citealt{plavin+2020, giommi+2020, plavin+2021, smith+2021, hovatta+2021, zhou+2021, bartos+2021, kun+2022, buson+2022, plavin+2023, novikova+2023, buson+2023, plavin+2023_xray_connection, bellenghi+2023_5bzcat_and_rfc_expansion, suray2024}).

Most of the aforementioned studies investigated the possible blazar-neutrino correlation in a purely spatial way. While the spatial information of the neutrino events is challenging to accurately determine, the temporal information of the neutrino arrival is very accurately recorded. These arrival times are needed to enhance the detection power of a correlation test to reliable levels by reducing the probability of chance associations (e.g. \citealt{liodakis+2022_wild_hunt}).

One way is to look for temporal associations between the neutrino arrival times and high flux states (typically referred to as "flaring" states) in the light curve of blazars spatially associated with them. This has the underlying assumption that neutrino flux increases as the jet becomes more active and the multiwavelength fluxes increase during flaring periods. For example, during X-ray and $\gamma$-ray flaring periods, it has been estimated that the neutrino flux could possibly reach the detectable range of the IceCube Neutrino Observatory (e.g. \citealt{oikonomou+2019_indepth_blz_neut_connection, kreter+2020_fermi_flare_v_neutrinos, stathopoulos+2022_xray_flare_v_neutrinos}). On the other hand, in lower energies (radio and optical), blazar flares can trace the global jet activity. However, due to longer particle cooling times, these lower energy flaring periods are longer in duration. Such longer-term increase in jet power may be relevant to the potential multi-messenger emission of blazar jets, as they could trigger current-driven kink instabilities and efficient particle acceleration (e.g. \citealt{Nalewajko2017_jet_power_kink_stability}).

Despite the aforementioned increase in detection power offered by utilizing the temporal information, \cite{liodakis+2022_wild_hunt} demonstrated that even if every neutrino event was accompanied by a blazar flare, due to the fundamental limitations of neutrino astronomy (low number of neutrinos with high likelihood of being of astrophysical origin and relatively large error bars in sky localization), the correlation significance between the blazars and neutrinos would be at best $\sim$3$\sigma$.

Due to blazar monitoring data being available only for a few wavelengths, most of the spatio-temporal studies have considered either individual or a handful of sources (e.g. \citealt{krauss+2014, kadler+2016, righi19, Franckowiak20}).
To bypass such sample size limitations, \cite{plavin+2020} performed a spatio-temporal analysis in the radio band using the RATAN-600 data (e.g. \citealt{kovalev+2002_ratan600_recent_paper}) of 1099 blazars, which resulted in a $\sim$2$\sigma$ blazar-neutrino spatio-temporal correlation. We note that \cite{plavin+2020} also found a $\sim$3$\sigma$ spatial-only blazar-neutrino correlation using a sample of 3388 flux-limited, VLBI-based blazars.
In \cite{hovatta+2021}, hereafter H21, we performed a spatio-temporal analysis using well-sampled Owens Valley Radio Observatory (OVRO) and Mets\"ahovi Radio Observatory light curves of 1795 blazars, which had significantly better cadence than the RATAN-600 light curves.
We showed that although not all neutrino events were associated with strong radio flaring blazars, when the associated neutrinos arrived during strong radio flares it had a chance probability of $\sim$2$\sigma$. 

In this paper, we perform similar spatio-temporal analyses on the connection between blazar flares and IceCube neutrinos. We start by revisiting the analysis done in H21; we increase the neutrino sample by three years (see \S \ref{sec:data_neut}) and the radio sample by two years from the Candidate Gamma-Ray Blazar Survey Source Catalog, CGRaBS, blazars (see \S \ref{sec:data_blz}) to see if the results remain consistent with our previous study or if we see an increase in the significance, like expected from our simulations in \cite{liodakis+2022_wild_hunt}. However, we note that after the publication of H21, the IceCube collaboration released a coherently selected catalog of HE neutrino events (\citealt{abbasi+2023_cat1}), rendering the neutrino sample used in H21 obsolete. Moreover, the cut-based neutrino selection of H21 has inherent caveats (see \S \ref{sec:data_neut}) which further reduce the reliability of the H21 neutrino sample. Therefore, in the main analysis of this paper, we utilize the most recent, coherently selected sample of IceCube neutrinos, and we upgrade our methodology with a weighting scheme that allows us to include all these neutrino events into the statistical analysis, without needing any selection thresholds (see \S \ref{sec:analysis_upgraded_spatiotemp}). Furthermore, we extend this upgraded spatio-temporal analysis to the optical regime as some of the weak radio sources show more variability in the optical band (e.g. \citealt{lindfors16}).

The paper is structured as follows:
In \S \ref{sec:data}, we describe the data used.
In \S \ref{sec:analysis}, the methodology of the spatio-temporal analysis as well as our upgraded version are presented.
In \S \ref{sec:results_and_disc}, the multi-band spatio-temporal results are presented and discussed.
In \S \ref{sec:conclusion} we provide concluding remarks.

\section{Data} \label{sec:data}
\subsection{Neutrino events} \label{sec:data_neut}
In this study we use the most recent HE neutrino events released by the IceCube collaboration operating the Antarctica-based 1 $\mathrm{km}^3$ IceCube Neutrino Observatory (see \citealt{abbasi+2023_cat1, abbasi+2022_extra_neut}). 
The complete list of the neutrino events used is given in the electronic table \texttt{Neut.dat}.

For the association analysis, it is crucial that the neutrino events have a good spatial localization and a high likelihood of being astrophysical.
Therefore, in H21, we collected muon track events that had 90\% confidence level error region size (on the celestial sphere) of less than 10 deg$^2$ and an estimated energy of at least 200 TeV.
With these limits we aimed to include only neutrinos with >50\% probability of having astrophysical origin.
The list included 56 events occurring between August 2009 and July 2019.
Since H21, the IceCube collaboration has made several major data releases (see below), and in this work, we use their most up to date neutrino lists.

IceCube started to send out alerts to the community on high signalness\footnote{
The chance that the event is of astrophysical origin by assuming an astrophysical neutrino spectral index of $2.19\pm0.10$ and calculating the ratio of the expected number of astrophysical neutrinos to the total (expected+background) number of neutrinos at a given event energy and declination (e.g. \citealp{abbasi+2023_plavin_test}). The value is given in IceCat-1 (\citealt{abbasi+2023_cat1}) for all neutrino events.} events in April 2016, with Bronze and Gold alerts corresponding to events with at least 30\% and 50\% signalness, respectively.
Recently, the first catalog of high signalness events was released (IceCat-1; \citealt{abbasi+2023_cat1}).
It includes a total of 275 events occurring between May 2011 and December 2022, eight of which we ignore due to being induced by cosmic rays.
For all events in IceCat-1, the neutrino arrival time, containment area on the celestial sphere, energy, and signalness are reported.
It should be noted that due to the starting date of May 2011 and the limitation to "online" analysis, some good quality muon track events are not reported in IceCat-1.
Therefore, we searched for other recent IceCube publications to obtain all the muon track events.

\cite{abbasi+2022_extra_neut} presented improved characterization of the astrophysical muon-neutrino flux with 9.5 years of IceCube data between May 2009 and December 2018.
It included a list of "offline" analyzed muon-neutrinos with the energy greater than 200 TeV and signalness larger than 50\%.
Due to the earlier start date and the use of "offline" analysis, there are 16 additional events in \cite{abbasi+2022_extra_neut} which are not present in IceCat-1.
We add these to our sample, bringing the total number to 283 neutrinos (hereafter referred to as IceCat1+ sample, see Figure \ref{fig:all_blz_neut_map}).
We note, however, that this selection is not completely homogeneous due to the differences in the cutoff energy and signalness limits.

We note that out of the 56 neutrino events included in H21, all except five events are present in IceCat1+.
These missing events have arrival dates of 2010-06-23 (MJD 55370.74), 2011-03-04 (MJD 55624.95), 2013-09-07 (MJD 56542.79 with the IceCat-1 ID of IC130907A* where it is marked as cosmic ray induced), 2016-03-31 (MJD 57478.60 with the IceCat-1 ID of IC160331A where it is present with a dramatically different RA), and 2019-03-31 (MJD 58573.16).
While many events are present in both IceCat1+ and the H21-selected neutrino sample, it is crucial to note that most of these shared events have different error regions and energies because of the improved event reconstruction algorithm. Many of the H21-selected neutrino events, now (in IceCat1+) have energies below the H21 energy threshold of 200~TeV. This means that the H21-selected neutrino sample greatly differs from IceCat1+ neutrino sample when the weighting scheme is employed (see \S \ref{sec:analysis_upgraded_spatiotemp}).

In the electronic table \texttt{Neut.dat} and through out this paper, the IceCat1+ neutrino events are referred to by their IC ID as given in IceCat-1 \citep{abbasi+2023_cat1}. When an event does not have such an ID assigned (the ones from \citealt{abbasi+2022_extra_neut}), we have generated one using its arrival MJD in the style of IceCat-1 with the difference that the letter assignment (the right-most element of the ID) for such cases starts from 'X' instead of 'A'. This is to distinguish the IDs generated by us from the official IceCube IDs.

\subsection{Blazar radio and optical light curves} \label{sec:data_blz}
\begin{figure*}
\centering
\includegraphics[width=18.5cm, keepaspectratio]{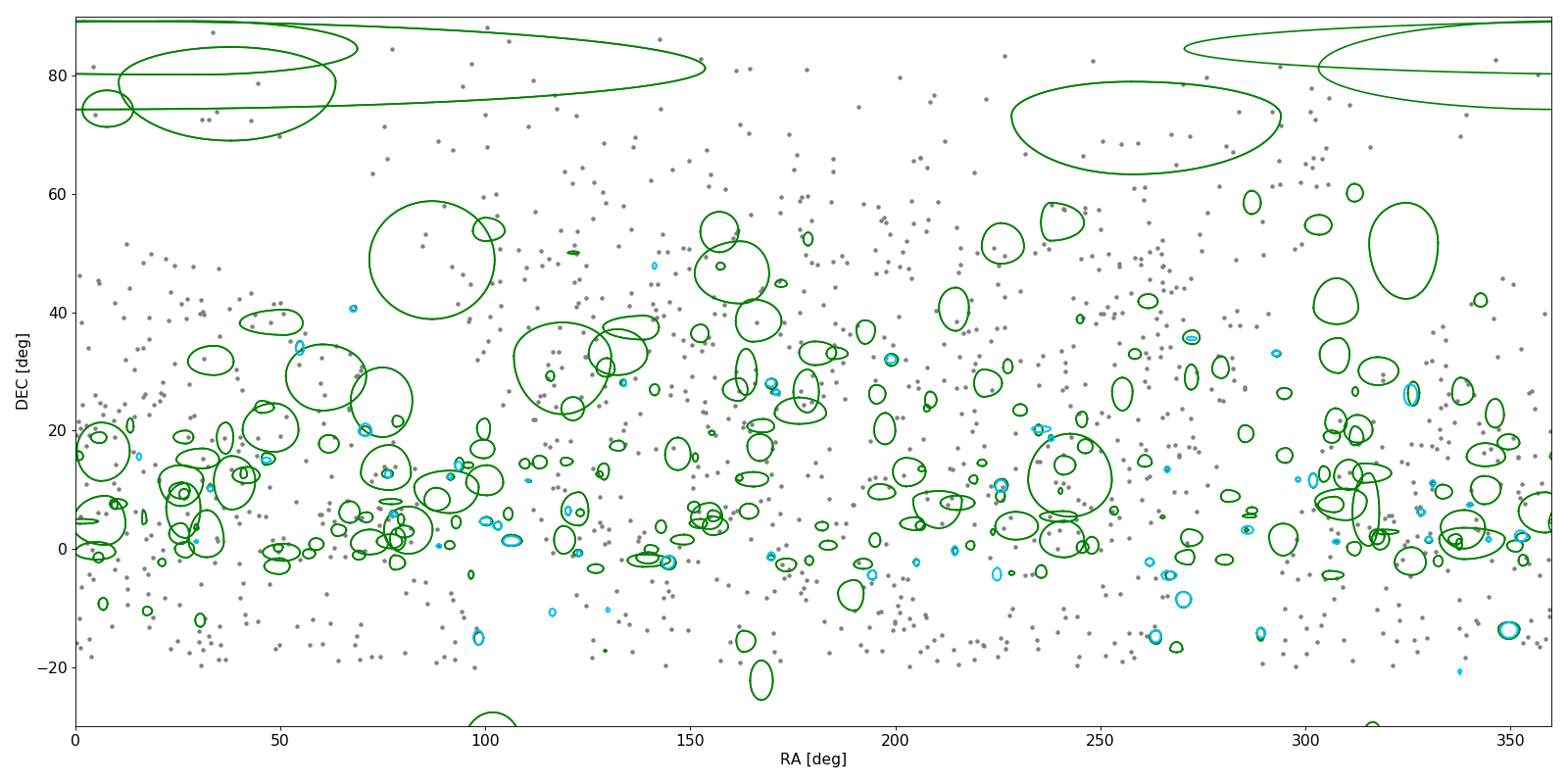}
\caption{The sky distribution of 277 IceCat1+ neutrinos (green ellipsoids), the 56 H21 neutrinos (blue ellipsoids), and 1157 CGRaBS blazars (grey dots). Six neutrino events are hidden due to lower than $-$30\degree\, declination. Additionally, all CGRaBS blazars have a declination greater than $-$20\degree\, due to the OVRO declination limit. The neutrino ellipses show the 90\% confidence level sky error regions as reported by IceCube. The full list of IceCat1+ neutrinos is given in the electronic table \texttt{Neut.dat}.}
\label{fig:all_blz_neut_map}
\end{figure*}

We use the blazar sample taken from the CGRaBS flat-spectrum radio quasar catalog \citep{healey+2008} which includes blazars monitored by the OVRO blazar monitoring program for over a decade \citep{richards+2011}.
The CGRaBS sample was the most statistically complete of the five blazar samples used in H21, which is why we opted to use it in this study.
As discussed in \citet{richards+2011, richards14}, the CGRaBS sample is complete to 65~mJy flux density at 4.8~GHz and radio spectral index $\alpha > -0.5$ where S $\propto \nu^{\alpha}$. However, the CGRaBS sample does not represent the full blazar population as it is dominated by low synchrotron peaked objects (LSPs) that are radio bright (see discussion in H21 and in \S \ref{sec:upgraded_results_radio_only}). Moreover, the CGRaBS sample excludes objects near the Galactic plane with Galactic latitude $ |b| < 10\degree$.
In total it includes 1157 blazars with OVRO radio light curves extending from 2008 to the end of 2022.
The sample does not include any blazars with a declination below $-$20\degree.
The blazar coordinates along with their common aliases are listed in the electronic table \texttt{Blz.dat}.
Figure \ref{fig:all_blz_neut_map} shows the sky distribution of these blazars as well as the IceCat1+ and H21 neutrinos (see \S \ref{sec:data_neut}).

In the optical regime, we combined CRTS\footnote{\href{https://crts.caltech.edu/}{https://crts.caltech.edu/}} \citep{drake+2009}, ATLAS\footnote{\href{https://fallingstar-data.com/}{https://fallingstar-data.com/}} \citep{tonry+2018}, and ZTF\footnote{\href{https://irsa.ipac.caltech.edu/Missions/ztf.html}{https://irsa.ipac.caltech.edu/Missions/ztf.html}} \citep{bellm+2019} optical light curves of 1061 CGRaBS blazars (for 96 blazars we did not find any optical data).
The light curves are obtained by performing forced-photometry on the aforementioned all-sky databases using the CGRaBS radio coordinates to an accuracy of 0.001\degree\, (see Appendix \ref{appendix:caz_combination}).
Additionally, whenever possible, KAIT\footnote{\href{https://w.astro.berkeley.edu/bait/kait.html}{https://w.astro.berkeley.edu/bait/kait.html}} \citep{filippenko+2001, liodakis+2018_cross_correlation1} and Tuorla\footnote{\href{https://users.utu.fi/kani/1m/}{https://users.utu.fi/kani/1m/}} \citep{nilsson+2018} optical light curves were added onto the combined all-sky light curves.

The reduction and combination procedures for the CAZ (CRTS+ATLAS+ZTF as well as KAIT and Tuorla when available) optical light curves are explained in detail in Appendix \ref{appendix:caz_combination}. The final CAZ light curves considered are limited to those that have at least 10 data points and are at least 30 days in total duration. The longest light curves have data from the beginning of the CRTS survey in 2008 up to 2022 (limited by the latest ZTF data release used in this work).

\section{Analysis} \label{sec:analysis}
In this paper, we look for potential spatio-temporal correlations between the radio and optical characteristics of CGRaBS blazar light curves and IceCube HE ($\gtrsim$100 TeV) neutrino events.
We use the methodology of H21 and \cite{plavin+2020} with the addition of taking neutrino signalness and error region size into consideration.
This upgraded spatio-temporal methodology is described in this section.

\subsection{The spatio-temporal method} \label{sec:analysis_original_spatiotemp}
The initial step in the spatio-temporal method is to find the blazars that are spatially correlated with the observed neutrino events. This is achieved by cross-correlating the blazar coordinates with the 90\% likelihood error regions of the neutrino events (sometimes referred to as the neutrino error regions in this paper). Spatially correlated blazars are those that fall within the neutrino error regions (i.e., the spatial correlation is treated as a Boolean condition of blazars either being in or out of a neutrino error region).

In the next step, a few test-statistic (TS) parameters are chosen based on the known observational properties of the blazars. In H21, the main observational property of interest was the blazar flux density (at 15~GHz)\footnote{See \S \ref{sec:intro} for motivation.}, which allowed for four independent TS parameters to be used in the statistics. In this study we use the same four TS parameters: mean 15~GHz flux density ($\overline{S}$), activity index (AI), and the number of associated blazars flaring with a 99\% and a 99.99\% confidence level at the time of the neutrino arrival (AI$_{1\%}$ and AI$_{0.01\%}$, respectively).

\begin{figure*}
\centering
\includegraphics[width=18.5cm, keepaspectratio]{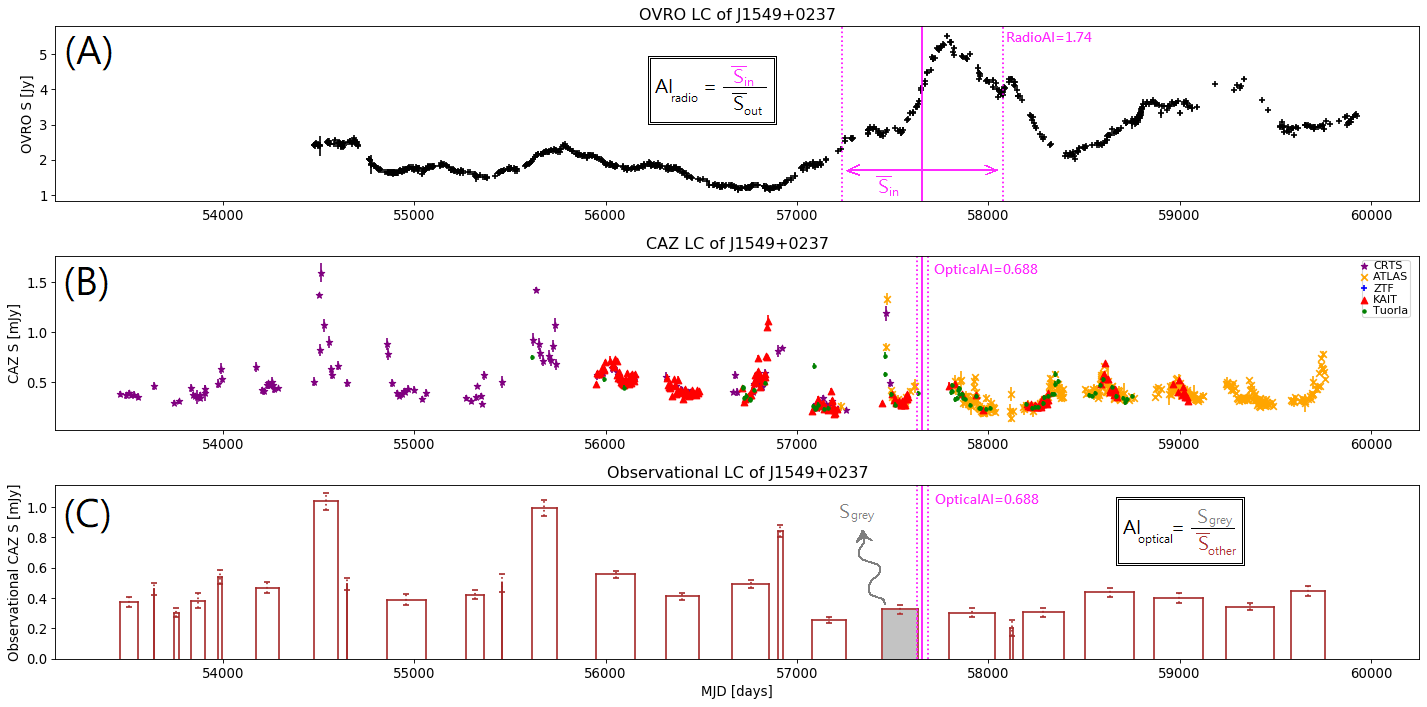}
\caption{The radio and optical light curves of the blazar J1549+0237 with a spatially-associated neutrino, IC160924A. Top panel (A) shows the radio (OVRO) light curve. Middle panel (B) shows the optical (CAZ) light curve. Bottom panel (C) shows the optical (CAZ) light curve when converted into observational blocks (see \S \ref{sec:analysis_blz_lc_optical_only}). The solid vertical (magenta) line shows the moment of neutrino arrival. In case of radio, the time window for AI calculation is 2.3 years (see \S \ref{sec:analysis}), shown by dotted vertical lines. In optical, the observational block used for AI calculation is within $\pm$30 days of the neutrino arrival, also marked using grey shading. For this association, the AI value in case of radio is 1.74 and in case of optical it is 0.688.}
\label{fig:eg_LC}
\end{figure*}

AI is the ratio of the mean flux density within a defined time window (centered at the neutrino arrival time) to the mean flux density outside the time window (see Figure \ref{fig:eg_LC}). When a time window in the light curve has AI>1, we define it as a "flaring" period. Due to the flux density errors of the data points, periods with AI>1 can occur due to random statistical fluctuations. Thus for the latter two TS parameters, based on the light curves in our sample, we determine\footnote{From our sample, we randomly select a light curve and randomize its flux density data points assuming Gaussian errors. Then we calculate a false-AI by taking the ratio of the mean flux density of the randomized light curve to the mean flux density of the original light curve. This process is repeated $10^5$ times and the resulting distribution is used to determine the 99\% and 99.99\% confidence thresholds.} confidence thresholds above which the enhanced AI value (AI>AI$_{\mathrm{threshold}}$>1) is unlikely to have occurred due to statistical fluctuations (i.e., the time window is confidently showing flaring behavior). We emphasize that the value of AI$_{\mathrm{threshold}}$ solely depends on the flux density errors of the data points. We note that while the AI in the radio and optical bands aim to quantify the same property (i.e, the average flux density level within a given time window relative to the historical level), they are calculated somewhat differently because of the presence of faster flaring-timescales and seasonal gaps in the optical light curves (see \S \ref{sec:analysis_blz_lc_radio_only} and \S \ref{sec:analysis_blz_lc_optical_only}).

For each of the four TS parameters ($\overline{S}$, AI, AI$_{1\%}$, and AI$_{0.01\%}$), a global TS value is calculated using all the spatially-associated blazars. In the case of $\overline{S}$ and AI, the global TS value is obtained by averaging the mean flux densities and AI values of each individual spatially-associated blazar globally, respectively. In the case of AI$_{1\%}$ and AI$_{0.01\%}$, the global TS value is obtained by counting the number of blazar-neutrino associations whose AI value exceeds the 99\% and 99.99\% AI thresholds, respectively. To obtain a null hypothesis (random) TS value distribution for each TS parameter, the RA of the neutrino events is randomized while keeping all other properties of the observed neutrino events the same (for example see \citealt{aartsen+2017c}). One randomized neutrino setup will result in one random TS value for each TS parameter. To obtain a distribution of such random TS values, we repeat the randomization a total of $10^5$ times.

\begin{figure*}
\centering
\includegraphics[width=18.5cm, keepaspectratio]{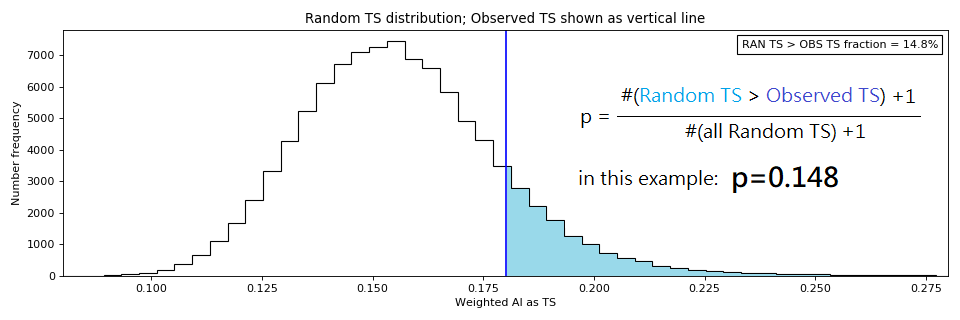}
\caption{The distribution of the TS values for $10^5$ random realizations is shown in black, along with the observed TS value shown as a blue vertical line. In this example, the TS parameter used is AI on the CGRaBS observational optical light curves (see \S \ref{sec:analysis_blz_lc_optical_only}). The region of the distribution shown in cyan is where random TS values are greater than the observed TS value, whose fraction gives the p-value to reject the null hypothesis (see Equation \ref{eqn:pval}).}
\label{fig:TS_eg_AI}
\end{figure*}

Using these random TS value distributions, the chance probability that the blazars and the observed neutrino events are uncorrelated (i.e., the p-value of rejecting the null hypothesis) for each TS parameter is found by counting the fraction of random TS values that are greater than the TS value in the observed setup:
\begin{equation} \label{eqn:pval}
    p = \frac{M+1}{N+1}
\end{equation}
where $p$ is the p-value, $M$ is the number of realizations where a random TS value is larger than the observed TS value, and $N$ is the total number of random realizations (\citealt{davison_hinkley_1997}). This is visualized in Figure \ref{fig:TS_eg_AI}. We apply these tests to three sets of CGRaBS light curves: the radio light curves, the optical light curves, and their combination.

\subsubsection{The radio-only analysis} \label{sec:analysis_blz_lc_radio_only}
To calculate the AI value in the radio band, as done in H21, we use a fixed time window of 2.3 years centered around the neutrino arrival time; see panel (A) of Figure \ref{fig:eg_LC}.
This duration was found to be typical for 15 GHz flares by \cite{hovatta07}.
The 1\% and 0.01\% AI false-positive values used to identify flaring blazars in the radio band are 1.10 and 1.29, respectively, as determined in H21.
We note that although some of the radio light curves occasionally have extreme outliers, we did not attempt to clean them because the 2.3 years time window is long enough to average out their influence on the statistics.

\subsubsection{The optical-only analysis} \label{sec:analysis_blz_lc_optical_only}
While the typical flaring timescales of blazar light curves in the radio band are on the order years (e.g. \citealt{hovatta07}); in the optical band, blazars typically exhibit faster variability due to the higher energy of the electrons producing the synchrotron radiation. The optical flaring timescales vary from hours, days, to months, and sometimes even longer (e.g. \citealt{Marscher2021_optical_intra_night_variability, Raiteri2017_intraday_var, lindfors16}). This behavior may be due to a complicated mixture of isolated particle acceleration events (fast flares) as well as global changes in accretion and/or jet parameters (months- to years-long flares). As a result, since the longer-term jet activity changes could be relevant to the multi-messenger emission of the jet (see \S \ref{sec:intro}), we favor longer-term optical flares over the shortest flares in this study.

Unfortunately, optical light curves exhibit seasonal gaps arising from inherent observational limitations that the radio band does not suffer from. Therefore, we could not calculate the optical AI values in the same way as the radio band, especially since a typical flaring timescale is complicated to quantify for such discontinuous light curves. Instead, we utilized the following data-limited approach.

For each optical light curve, we calculated the time difference between two consecutive data points. We then identified pairs where the time difference was longer than 30 to 60 days\footnote{The exact value depends on the distribution of the consecutive time gaps across a light curve and is the most frequent value between 30 and 60 days.} to identify observational gaps. Subsequently, we calculated the weighted average of the flux densities between two observational gaps to form observational blocks. For example, the middle panel of Figure \ref{fig:eg_LC} shows the original CAZ light curve of the blazar J1549+0237 and the bottom panel shows its corresponding observational light curve composed of observational blocks. We note that most, but not all, observational blocks correspond to observational seasons. Hereafter, an \textit{observational} light curve refers to the optical CAZ light curve converted into observational blocks.

To calculate the AI value in the optical regime, we searched for the closest observational block within $\pm$30 days to the neutrino arrival time and calculated the AI value by taking the flux density ratio of the closest block to the mean of all other observational block flux densities.
If a neutrino arrives in an observational gap and the nearest observational block is more than 30 days away, we omit its blazar-neutrino association from the optical AI-related tests.
An example of the AI calculation in the optical regime is shown in Figure \ref{fig:eg_LC} (panel C).
The $\pm$30-day leeway limit is chosen as it is the smallest temporal length that can fill the smallest of observational gaps (i.e., the limit that most conservatively minimizes the number of omitted AI-related tests). A larger limit would add too much uncertainty to the average flux density level at the moment of neutrino arrival, while a smaller limit would result in many "near miss" omissions. We recognize the caveat that, when a nearby observational block is on the order of days, a $\pm$30-day leeway may result in an uncertain flux level arising from only a few data points. However, given that such short observational blocks are rare, their effects are expected to average out.

Using the optical observational light curves we find that the 1\% and 0.01\% AI false-positive values used to identify flaring blazars in the optical band are 1.11 and 1.25, respectively. We emphasize that we use the observational light curves to obtain the optical AI values. As a result, the aforementioned "flaring" behavior corresponds to long-term, above-average optical activity. Such optical flares are distinct from short-term ones often discussed in the literature.

\subsubsection{The simultaneous radio+optical analysis} \label{sec:analysis_blz_lc_r+o}
To combine the light curve information of a specific blazar that has both radio and optical light curves, we normalized the radio and optical flux densities by their respective global mean values.
After the normalization we obtained, the TS value $\overline{S}$ of one radio+optical association by taking the average of the two normalized radio and optical mean flux densities.

If an optical AI value existed, we considered the radio+optical AI value as the average of the two individual radio and optical AI values.
Otherwise, if a radio+optical AI value could not be determined, we omitted the association.
Additionally, the radio+optical TS parameters AI$_{1\%}$ and AI$_{0.01\%}$ were found by counting the number of blazars for which the radio and optical AI values cross their respective false-positive thresholds simultaneously.

We note that while radio and optical light curves generally do not have comparable flaring timescales, observational optical light curves (with each observational block duration being on the order of several months) trace slower flux variations which are comparable to the year-long radio variations. This timescale comparability enables us to perform the aforementioned simultaneous radio+optical analysis. However, we recognize the caveat that the radio and optical time windows are not constructed identically, since the former time windows are dynamically fixed to the neutrino arrival time while the latter ones are observationally constrained.

\subsection{The upgraded spatio-temporal method} \label{sec:analysis_upgraded_spatiotemp}
In order for the spatio-temporal method to be able to detect a potential correlation between blazars and neutrinos, the effect of the events with low signalness and/or high directional uncertainty on the global TS needs to be diminished as their contribution is intrinsically mostly random noise.
In this subsection, we upgrade the spatio-temporal method by introducing a weighting scheme that diminishes the effect of such events gradually, which allows for all IceCat1+ neutrino events to be used in the analysis without needing to set arbitrary selection thresholds as has been done in previous studies (H21, \citealt{plavin+2020}).
While such pre-defined thresholds are more efficient at removing the noise contribution from the "inaccurate" events, the weighting scheme is less arbitrary, preserves the signal from the intermediate events (with parameters close to the thresholds), and is more resilient to slight parameter changes that could be introduced in the future modifications of the event reconstruction techniques.

In the weighting scheme, we assign a weight, $W$, to each neutrino event via Equation \ref{eqn:weight}. The weight is implemented into the statistics prior to the step when the global TS value for each of the TS parameters is calculated (see \S \ref{sec:analysis_original_spatiotemp}). For example, in the case of AI, the AI value of each spatially-associated blazar-neutrino pair is multiplied by $W$ of the corresponding neutrino event before calculating the global average. In other words, in the upgraded spatio-temporal method, the weight of each neutrino event affects how much each of their corresponding spatially-associated blazars contribute to the global TS values.

As IceCube provides the 90\% confidence level sky error regions for each event in the form of four asymmetrical ellipse regions in four quadrants, it is possible to calculate the error region size, $\Omega$, of each neutrino event via
\begin{equation} \label{eqn:omega}
    \Omega = \frac{\pi}{4}\left(\alpha^+\cdot\delta^++\alpha^-\cdot\delta^++\alpha^-\cdot\delta^-+\alpha^+\cdot\delta^-\right)
\end{equation}
where $\alpha^+$, $\alpha^-$, $\delta^+$, and $\delta^-$ are directional errors for the positive RA, negative RA, positive DEC, and negative DEC , respectively.
The weight, $W$, for each neutrino event is defined by combining its $\Omega$ and signalness, $\mathcal{S}$, via 
\begin{equation} \label{eqn:weight}
    W =
    \begin{cases}
      \mathcal{S} & \text{if $\Omega \le \widetilde{\Omega}$}\\
      \mathcal{S} \cdot \widetilde{\Omega} \ / \ \Omega & \text{if $\Omega > \widetilde{\Omega}$}
    \end{cases}
\end{equation}
where $\widetilde{\Omega}$ is the global median $\Omega$ of all the IceCat1+ neutrinos.
The way $\mathcal{S}$ is implemented into this weighting scheme is similar to how \cite{abbasi+2023_plavin_test} handled their statistics.

\begin{figure}
\centering
\includegraphics[width=9cm, keepaspectratio]{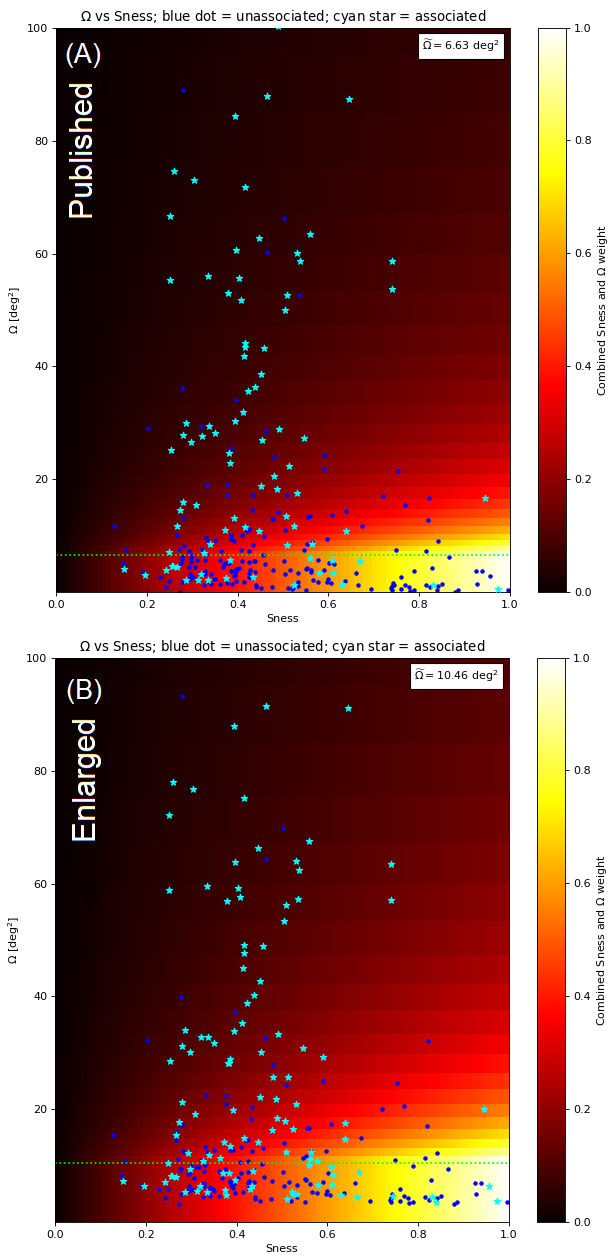}
\caption{The distribution of the IceCat1+ neutrino events in terms of the error region size ($\Omega$) and signalness ($\mathcal{S}$) is shown for: (A) the published, and (B) the enlarged error region scenarios (see \S \ref{sec:analysis_err_region}). The neutrino events that have at least one blazar association are shown as cyan stars while those that have no blazar association are shown as blue dots. The horizontal dotted line shows $\widetilde{\Omega}$ (median $\Omega$ of IceCat1+). The weight as calculated by Equation \ref{eqn:weight} is shown as a heat-map ranging from 0 to 1. Note that in these plots 13 neutrinos with $\Omega$>100 deg$^2$ are not shown.}
\label{fig:heatmaps}
\end{figure}

Equation \ref{eqn:weight} ensures that $W$ is always between 0 and 1 as $\Omega$ has no upper limit and $\mathcal{S}$ goes from 0 to 1.
In addition, when $\Omega \le \widetilde{\Omega}$, only $\mathcal{S}$ affects $W$.
This ensures that the global statistics is not overpowered by a handful of events that have the smallest $\Omega$.
The aim of this weighting scheme is to diminish the effect of random noise arising from the inclusion of the most poorly reconstructed events (small $S$ and/or large $\Omega$), without softening the potential signal that may come from the mid-range events (when $\mathcal{S}$ and $\Omega$ are close to their respective global medians, i.e., $\mathcal{S} \sim \widetilde{\mathcal{S}}$ and $\Omega \sim \widetilde{\Omega}$).
The heat-maps in Figure \ref{fig:heatmaps} show how $W$ varies with respect to $\Omega$ and $S$.

\subsection{Handling the unknown systematic errors of neutrinos} \label{sec:analysis_err_region}
While the IceCube collaboration attempts to account for all possible errors when estimating the size of the HE neutrino error regions, there may still exist some unknown systematic errors (e.g., \citealt{abbasi+2023_plavin_test}) arising due to ice inhomogeneities which cannot be fully accounted for.
\cite{aartsen+2013} estimated an upper limit of 1\degree\, for such a systematic error.
Although the work is ongoing on constraining this parameter \citep{abbasi+2021_syserr_related, abbasi+2021_fancy_method, abbasi+2022_extra_neut, icecube_collab2022_ngc1068}, the most recent ones seem to suggest that the error is most likely negligible \citep{icecube_collab2022_ngc1068, abbasi+2023_plavin_test}.

At the moment, as it is not possible to accurately account for such systematic errors, we approach this problem by considering two scenarios:
\begin{itemize}
    \item Published error region: in this scenario the error regions are as provided in IceCat-1 (i.e., 90\% confidence level sky error regions). 
    \item Enlarged error region: in this scenario we add 1.0\degree\, in quadrature to each of the four directional errors.
\end{itemize}
Similar analysis involving unknown systematic errors was presented in \cite{abbasi+2023_plavin_test}.
Since the choice of the error region scenario affects all neutrino events, the median error region size ($\widetilde{\Omega}$) used in Equation \ref{eqn:weight} depends on the chosen scenario.
In the published scenario it is 6.63 deg$^2$, while in the enlarged scenario it is 10.46 deg$^2$. 
Figure \ref{fig:heatmaps} shows the heat-maps for both of these scenarios.

\section{Results \& Discussion} \label{sec:results_and_disc}
In this section, we present and discuss the results of the spatio-temporal method.
In \S \ref{sec:results_recreating_h21} we recreate the H21 results with an updated list of neutrinos and radio light curves.
In \S \ref{sec:results_upgraded} we apply the upgraded spatio-temporal method (see \S \ref{sec:analysis}) to the updated radio light curves and the observational optical light curves.

\subsection{Recreating the results of H21} \label{sec:results_recreating_h21}
In this subsection, we repeat the H21 analysis for the explicit purpose of reassuring the reader that we remain consistent with our previous study and investigating how the correlation significances change when the neutrino dataset is updated (while still using the H21 neutrino cuts). In \cite{liodakis+2022_wild_hunt}, we showed that if the blazar-neutrino spatio-temporal correlation is real, then a spatio-temporal correlation analysis is expected to result in smaller p-values upon using more data. This is what we aim to test in this subsection. The following tests are referred to as "H21-based" tests.

To repeat the H21 analysis which used blazar radio light curves to investigate their spatio-temporal correlation with IceCube neutrinos,
we used the updated radio light curves and two distinct neutrino lists: firstly the H21 neutrino list and secondly an updated version of it.
We updated the H21 list such that its events remained as they were\footnote{In the new IceCube event reconstructions the localization errors as well as the energies have changed for many of these events and some events listed in H21 are not in IceCat1+ (see \S \ref{sec:data_neut}), but we keep the events as reported in H21. We note that in the new reconstruction many of these events actually have energies $<200$~TeV. For the upgraded spatio-temporal analysis we used the most up-to-date localization errors and signalness values.} but any new event from IceCat1+ that passed the H21 energy (E$\ge$200~TeV) and error region size ($\Omega$<10~deg$^2$) thresholds was added to the updated H21 list.
The H21 list had 56 events to which we added 22 events.

First we began by repeating the analysis made in H21 using the same neutrino list and the updated radio light curves: no weighting is employed and the neutrino error regions are calculated by multiplying the localization errors by 1.30\footnote{This was introduced by \cite{plavin+2020} and used in H21 to convert 1D error regions into 2D ones. In the upgraded spatio-temporal method we do not use this factor anymore as the IceCube errors are already given in 2D (see \citealt{abbasi+2023_plavin_test}).}, then adding 0.9\degree\, in quadrature.
We obtain 17 blazar-neutrino associations which result in the p-values 0.039, 0.010, 0.059, and 0.014 for the four TS parameters $\overline{S}$, AI, AI$_{1\%}$, and AI$_{0.01\%}$, respectively, with the corresponding TS values being 1.93, 1.19, 8, and 5.

It is apparent that the two additional years of radio observations do not affect the H21 p-values\footnote{See Table 4 and Table 6 of H21.} (0.027, 0.010, 0.072, and 0.005, with the corresponding TS values 2.02, 1.20, 8, and 6 for the same 17 blazar-neutrino associations), with the only exception being the case of AI$_{0.01\%}$ for which the p-value has tripled.
The reason for this change is the association of the blazar J0341+3352 with the neutrino IC150831A during whose arrival the AI value was 1.27 (close to the threshold of 1.29). 
The extension of the radio light curve slightly lowered the AI value of this association making it fall below the threshold because the flare continued, increasing the average flux density outside of the 2.3 year window. 
Since the number of flaring sources is small, adding or removing one association can dramatically affect the p-value.
Such dependence on rather stochastic light curve behavior is generally an inevitable caveat of using AI as a TS parameter, as was also discussed in H21.

Secondly, we repeated the H21 analysis using the updated H21 list (H21 events plus new ones from IceCat1+ that pass the H21 thresholds) which resulted in 27 blazar-neutrino associations with the p-values 0.054, 0.116, 0.357, and 0.097 (the corresponding TS values are 1.32, 1.08, 8, and 5).
The blazars of the 10 new associations, as a result of adding 22 new neutrino events (typically with mid-range signalness, $\mathcal{S} \sim \widetilde{\mathcal{S}}$), are all rather radio-dim (when $\overline{S}$ of a source is smaller than the median of all $\overline{S}$) with four being non-variable.
Additionally, none of the new associations is flaring in radio (i.e., AI$\lesssim$1).
These associations caused a unanimous increase in all of the observed p-values, lowering the correlation significances of $\overline{S}$, AI, and AI$_{0.01\%}$ to below 2$\sigma$.

In H21, we had $\sim$10 years of data to which, in these H21-based tests, we added two years of radio data and three years of neutrino data. We note that the selection criteria of the neutrino events in H21 was rather suboptimal (in hindsight of IceCat-1, see \S \ref{sec:data_neut}). Additionally, the update to the neutrino data constituted only 22 new events with mid-range signalness. Therefore, it is highly likely that the update is neither extensive nor systematic enough to allow for the effect predicted by \cite{liodakis+2022_wild_hunt} to be reliably seen. As a result, we refrain from making any conclusions based on these comparative results.

\subsection{The upgraded spatio-temporal results} \label{sec:results_upgraded}
The upgraded spatio-temporal method, utilizing the weighting scheme for the neutrino signalness and error region size, is applied to both of the published and enlarged error region scenarios (see \S \ref{sec:analysis_err_region}).
The final unweighted and weighted TS values as well as the corresponding p-values of the weighted TS values are given in Table \ref{tab:final_pvals}.

\begin{figure*}
\centering
\includegraphics[width=18.5cm, keepaspectratio]{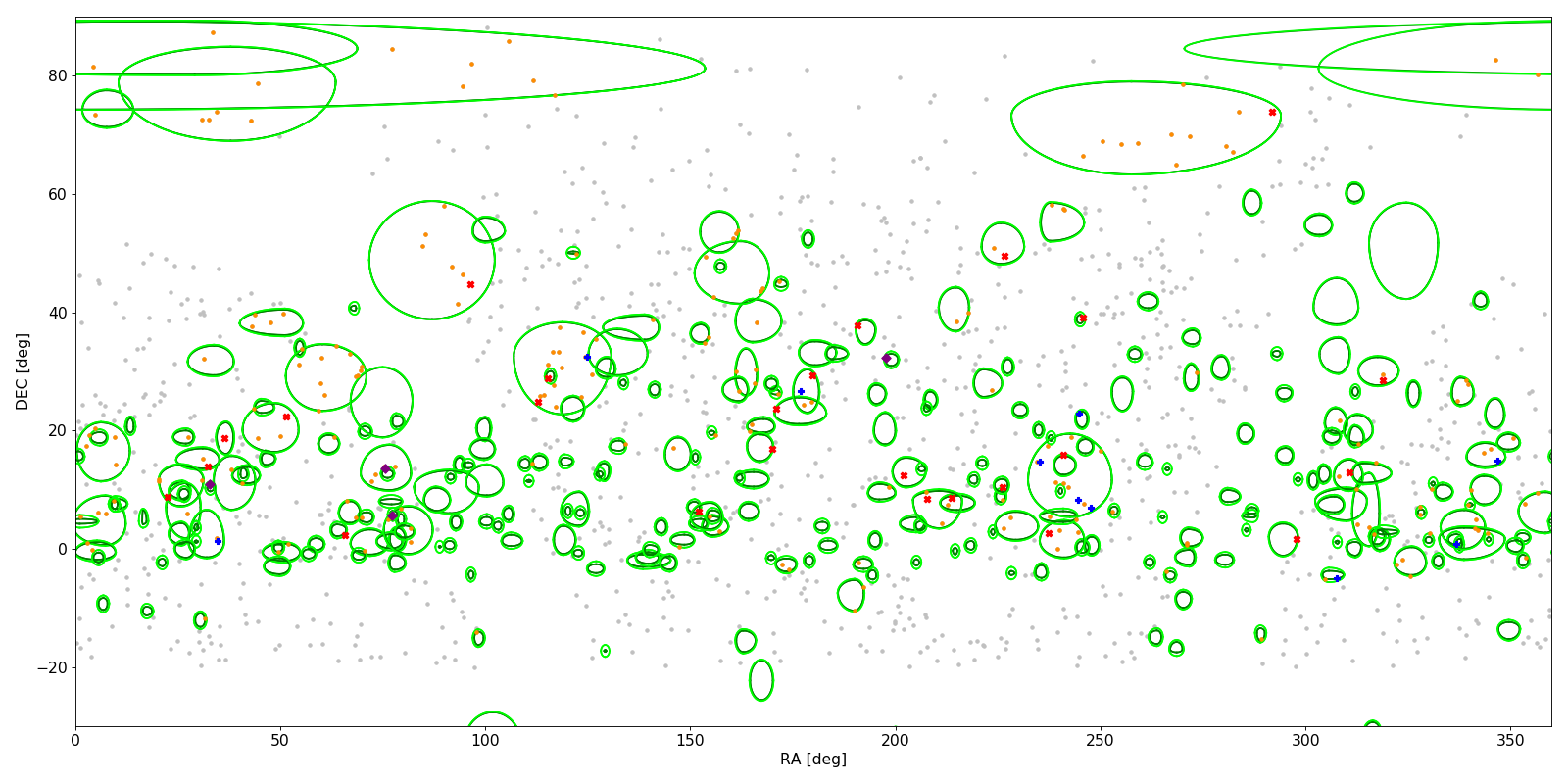}
\caption{The sky distribution of 277 IceCat1+ neutrinos (6 are below $-$30\degree\, DEC) and 1157 CGRaBS blazars. This plot is similar to Figure \ref{fig:all_blz_neut_map} with the exception that the published (dark green) and enlarged (light green) error regions are shown; note that by construction (see \S \ref{sec:analysis_err_region}) the larger the event region size, the smaller the relative difference between the published and enlarged error regions. When considering the enlarged error region scenario, there are 242 blazars with at least one flaring association. The grey dots (915) show unassociated blazars while the orange dots (203) show spatially associated blazars which have no flaring association. The red crosses (25) are blazars with at least one flaring association in radio and the blue plus symbols (10) are those with at least one flaring association in optical. The ones that show at least one simultaneous radio+optical flare are shown as purple diamonds (4:
J0502+1338--IC151114A,
J0211+1051--IC131014A,
J1310+3220--IC120515A, and
J0509+0541--IC190317A, where the last one is the second neutrino association with TXS~0506+056 which was not as spatially accurate as the one in 2017).}
\label{fig:flaring_neut}
\end{figure*}

The upgraded spatio-temporal method resulted in 275 and 231 blazar-neutrino associations in the enlarged and published error region scenarios, respectively, all of which are given in the electronic table \texttt{Assoc.dat}.
See Figure \ref{fig:flaring_neut} for a spatial visualization of the associations.
For comparison, H21 had only 17 associations.
This drastic difference in the number of associated sources arises because in the upgraded method all neutrino events are taken into consideration while in H21 no neutrino event with error region size larger than 10 deg$^2$ was considered.
For example, there are 38 events with $\Omega>50$ deg$^2$ which are responsible for 154 and 146 blazar-neutrino associations in the enlarged and published error region scenarios, respectively.
Additionally, the larger the positional error region of the neutrino event, the larger the chances of it randomly having multiple flaring associations (only one can be real).
The weighting scheme we introduced in \S \ref{sec:analysis_upgraded_spatiotemp} is designed to weigh down the contribution from such events; for example, as seen in Figure \ref{fig:flaring_neut} at around the coordinates (240, 15), the neutrino event IC200410A has multiple flaring associations but its contribution to the final statistics is negligible due to a small weight of 0.014.

We found that the CGRaBS blazars and IceCat1+ neutrinos are not correlated in the published error region scenario (i.e., when using the error regions reported in the IceCube data release).
Using the enlarged error regions, we found that the correlation with most of the spatio-temporal TS parameters (AI, AI$_{1\%}$, AI$_{0.01\%}$, and AI$_\mathrm{R+O}$) strengthens; a similar result was found in \cite{abbasi+2023_plavin_test}.
When taking the individual p-values into account, five p-values have higher significance than $2\sigma$ and, one of them, has higher significance than 3$\sigma$ (see Table \ref{tab:final_pvals}).
For a discussion on the trial-correction of these individual p-values see \S \ref{sec:trial_correction}.
Table \ref{tab:flr_assoc} shows flaring blazar-neutrino associations (99.99\% confidence level) arranged in descending order using their neutrino weights which determine their contribution to the observed AI$_{0.01\%}$ TS value (the TS parameter that gives the strongest correlations in general).
All the light curves of the aforementioned flaring blazars are given in Appendix \ref{appendix:all_flr_LCs}.

We emphasize that the updated H21 results discussed in \S \ref{sec:results_recreating_h21} are not directly comparable to the results of the upgraded method obtained in this section.
Apart from the obvious difference in methodology (cuts versus weights), in the updated H21 method the majority of neutrino events have outdated spatial information and differently constructed error regions.
These differences result in, for example, the bright blazar J1256-0547 (aka 3C~279, with $\overline{S}\sim$18 Jy) to be an association in the updated H21 method but not in the upgraded method.
J1256-0547 is solely responsible for the dramatically different $\overline{S}$ p-values (0.05 for updated H21 and 0.57 for the upgraded method).

\begin{table*} 
\caption{The unweighted and weighted TS values of the observed blazar-neutrino associations as well as the respective weighted p-values in the upgraded spatio-temporal method.}
\centering
\begin{tabular}{c|lcccc}
\hline
\hline
Error & Band & $\overline{S}$ & AI & AI$_{1\%}$ & AI$_{0.01\%}$ \\
(1) & (2) & (3) & (4) & (5) & (6) \\
\hline
\multirow{3}{*}{Published} & R & 0.58 $\rightarrow$ 0.05 (p=0.5719) & 0.99 $\rightarrow$ 0.10 (p=0.2252) & 54 $\rightarrow$ 4.91 (p=0.7739) & 22 $\rightarrow$ 2.53 (p=0.2963) \\
& O & 0.63 $\rightarrow$ 0.05 (p=0.3462) & 0.96 $\rightarrow$ 0.10 (p=0.4368) & 22 $\rightarrow$ 2.47 (p=0.4687) & 11 $\rightarrow$ 1.38 (p=0.4446) \\
& R+O & 1.07 $\rightarrow$ 0.09 (p=0.4491) & 1.00 $\rightarrow$ 0.11 (p=0.3372) & 7 $\rightarrow$ 0.419 (p=0.7580) & 1 $\rightarrow$ 0.023 (p=0.7588) \\
\hline
\multirow{3}{*}{Enlarged} & R & 0.58 $\rightarrow$ 0.09 (p=0.5445) & 0.99 $\rightarrow$ 0.16 (p=0.1652) & 66 $\rightarrow$ 11.7 (p=0.3149) & 29 $\rightarrow$ 7.10 ({\bf p=0.0107}) \\
& O & 0.59 $\rightarrow$ 0.08 (p=0.3595) & 0.99 $\rightarrow$ 0.18 (p=0.1477) & 31 $\rightarrow$ 7.01 ({\bf p=0.0352}) & 15 $\rightarrow$ 4.34 ({\bf p=0.0377}) \\
& R+O & 1.03 $\rightarrow$ 0.14 (p=0.4326) & 1.01 $\rightarrow$ 0.18 (p=0.1186) & 11 $\rightarrow$ 3.14 ({\bf p=0.0256}) & 4 $\rightarrow$ 2.270 ({\bf p=0.0014}) \\
\hline
\end{tabular}
\tablefoot{Column (1) gives the error region scenario (see \S \ref{sec:analysis_err_region}). Column (2) gives the blazar selection (R: radio-only, O: optical-only, R+O: simultaneous radio+optical). Columns (3) to (6) give the corresponding TS values (unweighted $\rightarrow$ weighted) as well as the weighted p-value for the TS parameters $\overline{S}$, AI, AI$_{1\%}$, and AI$_{0.01\%}$, respectively. The p-values shown in bold individually reach a significance level of >$2\sigma$. For a discussion on the trial correction of these individual p-values see \S \ref{sec:trial_correction}.}
\label{tab:final_pvals}
\end{table*}

\subsubsection{The radio-only results} \label{sec:upgraded_results_radio_only}
According to the p-values given in Table \ref{tab:final_pvals}, in the published error region scenario CGRaBS blazars and IceCat1+ neutrinos are not correlated.
In the enlarged error region scenario, the correlations strengthen.
However, only in case of using AI$_{0.01\%}$ as TS parameter does the correlation significance surpass 2$\sigma$.
This result is consistent with H21 in the sense that the highest correlation significance is found for the most confidently flaring blazars.

In case of $\overline{S}$ as TS parameter we do not find a significant correlation, which is in contrast to \cite{plavin+2023} who find that, on average, the brightest radio blazars from the RFC catalog\footnote{\texttt{http://astrogeo.org/rfc/}} have a >3$\sigma$ spatial correlation significance with IceCube HE neutrinos.
One reason for this discrepancy, as also discussed in H21, is that their blazar sample size (3412) is around three times larger than ours (1157) and not as dominated by low synchrotron peaked blazars (LSPs).
\cite{padovani+2016_HSPs_as_neut_emitters} suggested that high synchrotron peaked blazars (HSPs) are the neutrino emitting subclass of blazars which, if true, would mean that our sample is missing many neutrino emitting blazars. 

Additionally, the light curves used in this work are obtained from the OVRO monitoring program which uses a single-dish radio telescope including both the parsec-scale jet and the jet lobes in the beam.
Whereas the mean flux densities used by \cite{plavin+2023} are taken with Very-Long-Baseline Interferometry (VLBI) observations which contain no flux contribution from the jet lobes.
Due to this, the blazars in our sample are on average brighter than those of the RFC catalog.
Moreover, although the emission from the blazars in the CGRaBS sample is beamed, some radio galaxies (whose emission is not highly beamed)\footnote{According to \cite{hovatta+2009_agn_pop_properties}, the viewing angle of blazars is typically less than 10\degree\, while that of radio galaxies is greater than 10\degree.} were also included in the OVRO data set.
These radio galaxies are typically much brighter than the rest of the blazars due to a larger lobe contribution.
This means that the brightest sources in our sample are not the most beamed ones contrary to the blazar sample used by \cite{plavin+2023}.

The number of neutrino-associated blazars that are flaring in the radio band is 29 and 22 using the enlarged and published error regions, respectively.
These associations are listed in the top panel of Table \ref{tab:flr_assoc}.
We found that four of the five most confidently (99.99\% confidence level) flaring associations present in H21 are also flaring associations in the upgraded spatio-temporal analysis, namely in case of the blazars J1310+3220, J0211+1051, J0502+1338, and J1504+1029.
The fifth one (J0630-1323--IC170321A) is missing because it is no longer considered an association due to the neutrino event having slightly altered spatial parameters and systematic errors (see \S \ref{sec:results_recreating_h21}).

\subsubsection{The optical-only results} \label{sec:upgraded_results_optical_only}
We did not find significant correlations between the CGRaBS blazars in the optical band and IceCat1+ neutrinos when considering the published error region scenario.
Similar to the case of radio data, in the enlarged error region scenario, the correlations are stronger.
However, only for the confidently flaring sources (using AI$_{1\%}$ and AI$_{0.01\%}$ as TS parameters) we find that the individual correlation significances surpass 2$\sigma$.

It should be noted that interpreting $\overline{S}$ in the optical band is not straightforward.
The observed flux does not only arise from the parsec-scale jet, but also from the thermal components of the host galaxy and the big blue bump.
In general, the contribution from different components of a blazar is highly class- and redshift-dependent.
Therefore, it is not surprising that we do not see a correlation between the optical $\overline{S}$ and neutrinos.
However, interpreting the results of the AI-related TS parameters is more straightforward as flux density variability is only linked to temporal fluctuations of the jet emission as the other components are constant or show little variability.

Indeed, when comparing the radio-only and optical-only results, the correlation significances are highest when considering the blazar variability, especially those sources that were strongly flaring at the time of the neutrino arrival.
However, it should be noted that there are fewer optical than radio light curves (1061 versus 1157) and the optical light curves have observational gaps.
There are nine associations that were strongly flaring in radio but lack optical AI information due to an observational gap or lack of coverage in the neutrino arrival time period.
In other words, not all neutrino-associated blazars have an optical AI value which is why the number of optical associations is lower.
This is evident when comparing the top and bottom panels of Table \ref{tab:flr_assoc} which respectively show the radio and optical associations in case of AI$_{0.01\%}$ as TS parameter.

\subsubsection{The simultaneous radio+optical results} \label{sec:upgraded_results_simultaneous}
When repeating the upgraded spatio-temporal analysis on the radio and optical bands simultaneously, we once again do not find significant correlations between the CGRaBS blazars and IceCube neutrinos in the published error region scenario.
On the other hand, in the enlarged error region scenario, we find that the individual correlation significances surpass 2$\sigma$ and 3$\sigma$ when using AI$_{1\%}$ and AI$_{0.01\%}$ as TS parameters, respectively.

We note that out of the 283 IceCat1+ neutrinos, only 28 and 13 events are associated with the most confidently (99.99\% confidence level) flaring blazars at the time of their arrival in the radio and optical bands, respectively (see Table \ref{tab:flr_assoc}).
Therefore, it is clear that not all neutrinos originate from flaring blazars as also discussed in H21.

\subsubsection{Trial correction for the look-elsewhere effect} \label{sec:trial_correction}
We performed our spatio-temporal analysis for two statistically dependent error region scenarios (see \S \ref{sec:analysis_err_region}). For each of them, we obtained 12 individual p-values (see Table \ref{tab:final_pvals}). To account for the look-elsewhere effect of having checked these 24 individual p-values, we calculate the harmonic mean p-value ($\mathring{p}$, \citealt{Wilson2019_harmonic_mean_pval}) as follows:

\begin{equation} \label{eqn:hmp}
    \mathring{p} = \frac{\sum_{i=1}^{L} w_i}{\sum_{i=1}^{L} w_i/p_i}
\end{equation}
where $L$ represents the total number of p-values (24 in our case), $p_i$ represents each individual p-value, and $w_i$ represents the respective weight of each p-value such that $\sum_{i=1}^{L} w_i = 1$.  Under the general assumption that all the tested null-hypotheses are equally powerful in establishing a possible blazar-neutrino correlation, we obtain $w_i = 1/L = 1/24$. As a result, for the 24 tested p-values, Equation \ref{eqn:hmp} gives $\mathring{p} \approx 0.0250$.

In order to reject the null-hypothesis at a predefined threshold (with a false-positive rate of $\alpha$) using $\mathring{p}$, the predefined threshold $\alpha$ must be adjusted to take into account $L$ (i.e., the total number of null-hypotheses tested). We take the predefined threshold of 2$\sigma$, which has a false-positive rate of $\alpha_{2\sigma} = 0.0455$. To calculate the adjusted $\alpha_{2\sigma | L=24}$, we use Equation 4 of \cite{Wilson2019_harmonic_mean_pval}. This results in $\alpha_{2\sigma | L=24} \approx 0.0355$. Since $\mathring{p} \approx 0.0250 < 0.0355 \approx \alpha_{2\sigma | L=24}$, we conclude that $\mathring{p}$ satisfies the 2$\sigma$ threshold.

Additionally, we calculate the asymptotically exact harmonic mean p-value (via Equation 4 of \citealt{Wilson2019_harmonic_mean_pval}) of the 24 p-values with equal weights, which results in $p_{\mathring{p}} = 0.0299$. This can be directly compared to $\alpha_{2\sigma} = 0.0455$ and is equivalent to a false-positive rate of 2.17$\sigma$. Therefore, we conclude that the strength of the spatio-temporal correlation between the multiwavelength light curves of the CGRaBS blazars and the IceCube HE neutrinos, in the enlarged error region scenario, is at least 2.17$\sigma$.

\begin{table*}
\caption{The list of the neutrino-associated blazars that exhibit flaring behavior (99.99\% confidence level) in the enlarged error region scenario.}
\label{tab:flr_assoc}      
\centering          
\begin{tabular}{l|lcccccccccc}
\hline\hline
Band & Blazar name & Neutrino ID & T & $W$ & D$_\mathrm{BN}$ & D$_\mathrm{published}$ & D$_\mathrm{enlarged}$ & AI$_\mathrm{R}$ & wAI$_\mathrm{R}$  & AI$_\mathrm{O}$ & wAI$_\mathrm{O}$ \\ 
(1) & (2) & (3) & (4) & (5) & (6) & (7) & (8) & (9) & (10) & (11) & (12) \\ 
\hline 
\\
R+O & J0502+1338 & IC151114A & B & 0.957 & 1.06* & 0.72 & 1.27 & 1.61 & 1.54 & 1.52 & 1.46 \\ 
R & J1504+1029 & IC190730A & Q & 0.670 & 0.28 & 1.18 & 1.55 & 1.86 & 1.25 & 0.94 & 0.63 \\ 
R+O & J0211+1051 & IC131014A & B & 0.665 & 0.59* & 0.43 & 1.09 & 1.87 & 1.25 & 1.47 & 0.98 \\ 
R+O & J1310+3220 & IC120515A & U & 0.613 & 1.37* & 1.36 & 1.70 & 1.42 & 0.87 & 1.29 & 0.79 \\ 
R & J1623+3909 & IC190201A & Q & 0.533 & 0.80* & 0.75 & 1.25 & 1.31 & 0.70 & 1.09 & 0.58 \\ 
R & J0225+1846 & IC111216A & Q & 0.494 & 0.48 & 2.26 & 2.47 & 1.80 & 0.89 & 1.05 & 0.52 \\ 
R & J0422+0219 & IC141012A & Q & 0.436 & 2.07* & 1.77 & 2.13 & 1.29 & 0.56 & 0.92 & 0.40 \\ 
R & J1008+0621 & IC110726A & B & 0.396 & 1.12* & 1.03 & 1.46 & 1.52 & 0.60 & -- & -- \\ 
R & J1242+3750 & IC161001A & Q & 0.379 & 1.99* & 1.96 & 2.21 & 1.30 & 0.49 & -- & -- \\ 
R & J0740+2852 & IC200117A & Q & 0.375 & 0.93 & 1.12 & 1.52 & 1.34 & 0.50 & 1.07 & 0.40 \\ 
R & J1119+1656 & IC190422A & Q & 0.209 & 3.10 & 3.19 & 3.35 & 1.54 & 0.32 & 1.13 & 0.24 \\ 
R & J1351+0830 & IC180526X & Q & 0.122 & 2.86 & 5.99 & 6.17 & 1.50 & 0.18 & 1.09 & 0.13 \\ 
R & J1415+0832 & IC180526X & Q & 0.122 & 3.13 & 5.32 & 5.48 & 1.31 & 0.16 & 0.46 & 0.06 \\ 
R & J1327+1223 & IC140213A & - & 0.122 & 0.91 & 2.77 & 2.95 & 1.39 & 0.17 & -- & -- \\ 
R & J2114+2832 & IC150812A & Q & 0.114 & 1.92 & 2.86 & 3.07 & 1.46 & 0.17 & 1.06 & 0.12 \\ 
R & J1159+2914 & IC170527A & Q & 0.114 & 3.04 & 3.54 & 3.69 & 1.34 & 0.15 & 0.74 & 0.08 \\ 
R & J1951+0134 & IC171028A & Q & 0.108 & 3.41 & 3.55 & 3.69 & 1.32 & 0.14 & -- & -- \\ 
R & J0209+1352 & IC150428A & Q & 0.102 & 1.76 & 2.01 & 2.39 & 1.33 & 0.14 & -- & -- \\ 
R & J1549+0237 & IC160924A & Q & 0.094 & 3.97 & 5.36 & 5.47 & 1.74 & 0.16 & 0.69 & 0.06 \\ 
R & J2043+1255 & IC190410A & Q & 0.094 & 0.72 & 2.86 & 3.03 & 1.40 & 0.13 & 0.82 & 0.08 \\ 
R & J0130+0842 & IC181120A & Q & 0.090 & 4.32 & 4.85 & 4.95 & 1.33 & 0.12 & 0.86 & 0.08 \\ 
R & J1124+2336 & IC120529A & Q & 0.089 & 5.52 & 5.73 & 5.84 & 1.67 & 0.15 & 1.24 & 0.11 \\ 
R & J1506+4933 & IC140324A & Q & 0.071 & 1.79 & 3.26 & 3.44 & 1.41 & 0.10 & -- & -- \\ 
R & J0325+2224 & IC191231A & Q & 0.053 & 3.72 & 5.25 & 5.35 & 1.35 & 0.07 & 1.02 & 0.05 \\ 
R+O & J0509+0541 & IC190317A & B & 0.035 & 4.62 & 5.09 & 5.20 & 2.46 & 0.09 & 1.36 & 0.05 \\ 
R & J0731+2451 & IC140223A & Q & 0.015 & 9.74 & 10.46 & 10.51 & 1.38 & 0.02 & -- & -- \\ 
R & J1603+1554 & IC200410A & U & 0.014 & 4.60 & 8.05 & 8.11 & 1.35 & 0.02 & 0.72 & 0.01 \\ 
R & J0625+4440 & IC160812A & B & 0.011 & 10.22 & 13.83 & 13.87 & 1.42 & 0.02 & -- & -- \\ 
R & J1927+7358 & IC161127A & Q & 0.006 & 34.41 & 36.17 & 36.19 & 1.39 & 0.01 & -- & -- \\ 
\\
\hline
\\
R+O & J0502+1338 & IC151114A & B & 0.957 & 1.06* & 0.72 & 1.27 & 1.61 & 1.54 & 1.52 & 1.46 \\ 
R+O & J0211+1051 & IC131014A & B & 0.665 & 0.59* & 0.43 & 1.09 & 1.87 & 1.25 & 1.47 & 0.98 \\ 
O & J0509+0541 & IC170922A & B & 0.631 & 0.12 & 0.47 & 1.13 & 0.87 & 0.55 & 1.42 & 0.89 \\ 
R+O & J1310+3220 & IC120515A & U & 0.613 & 1.37* & 1.36 & 1.70 & 1.42 & 0.87 & 1.29 & 0.79 \\ 
O & J2030-0503 & IC180417A & Q & 0.559 & 1.97* & 1.97 & 2.75 & 0.88 & 0.49 & 2.33 & 1.30 \\ 
O & J1619+2247 & IC190413B & Q & 0.383 & 1.11 & 1.32 & 1.66 & 0.94 & 0.36 & 1.38 & 0.53 \\ 
O & J2307+1450 & IC120523A & B & 0.154 & 3.18 & 4.19 & 4.47 & 1.09 & 0.17 & 1.81 & 0.28 \\ 
O & J0219+0120 & IC161011X & Q & 0.136 & 2.71 & 4.17 & 4.29 & 1.05 & 0.14 & 1.29 & 0.17 \\ 
O & J1147+2635 & IC170527A & Q & 0.114 & 1.59 & 3.47 & 3.61 & 0.78 & 0.09 & 1.30 & 0.15 \\ 
O & J2226+0052 & IC200523A & Q & 0.036 & 2.14 & 5.22 & 5.34 & 0.97 & 0.04 & 1.47 & 0.05 \\ 
R+O & J0509+0541 & IC190317A & B & 0.035 & 4.62 & 5.09 & 5.20 & 2.46 & 0.09 & 1.36 & 0.05 \\ 
O & J0819+3226 & IC140223A & Q & 0.015 & 5.93 & 11.87 & 11.91 & 0.51 & 0.01 & 1.39 & 0.02 \\ 
O & J1630+0701 & IC200410A & Q & 0.014 & 6.86 & 7.67 & 7.75 & 1.18 & 0.02 & 1.59 & 0.02 \\ 
O & J1540+1447 & IC200410A & B & 0.014 & 8.03 & 9.68 & 9.74 & 0.67 & 0.01 & 1.36 & 0.02 \\ 
O & J1618+0819 & IC200410A & Q & 0.014 & 3.86 & 6.82 & 6.90 & 1.20 & 0.02 & 1.63 & 0.02 \\   
\end{tabular}
\tablefoot{ Column (1) gives the band in which the neutrino-associated blazar is detected to be flaring: R denotes flaring in radio only (unweighted radio AI>1.29), O denotes flaring in optical only (unweighted optical AI>1.25), and R+O denotes flaring in radio and optical simultaneously. Note that there are two main panels: the top panel shows the neutrino-associated blazars that are detected to be flaring in the radio band, while the bottom panel shows the neutrino-associated blazars that are detected to be flaring in the optical band; four neutrino associations occur simultaneously in both. Column (2) gives the neutrino-associated blazar CGRaBS name. Column (3) gives the IC ID of the associated neutrino (see \S \ref{sec:data_neut}). Column (4) gives the blazar 5BZCat type (T), if one exists; "B" denotes BL Lac, "Q" denotes flat spectrum radio quasar, and "U" denotes uncertain. Column (5) gives the neutrino weight ($W$) as calculated in the enlarged error region scenario. Column (6) gives the blazar angular distance to the associated neutrino (D$_\mathrm{BN}$); the symbol "*" indicates that the association only occurs in the enlarged error region scenario (i.e., D$_\mathrm{BN}$ > D$_\mathrm{published}$). Columns (7) and (8) give the angular distance between the neutrino and the error region edge (the 90\% confidence level limit) in the published and enlarged error region scenarios, respectively, along the blazar-neutrino vector. Note that all angular distances are given in degrees. Columns (9) to (12) give the unweighted AI value in radio (AI$_\mathrm{R}$), the weighted AI value in radio (wAI$_\mathrm{R}$) using enlarged errors, the unweighted AI value in optical (AI$_\mathrm{O}$), and the weighted AI value in optical (wAI$_\mathrm{O}$) using enlarged errors, respectively. In case of the optical band, in eight neutrino associations the neutrino arrives in an observational gap in which case AI$_\mathrm{O}$ and wAI$_\mathrm{O}$ are shown as "--".}
\end{table*}

\begin{figure*}
\centering
\includegraphics[width=18cm, keepaspectratio]{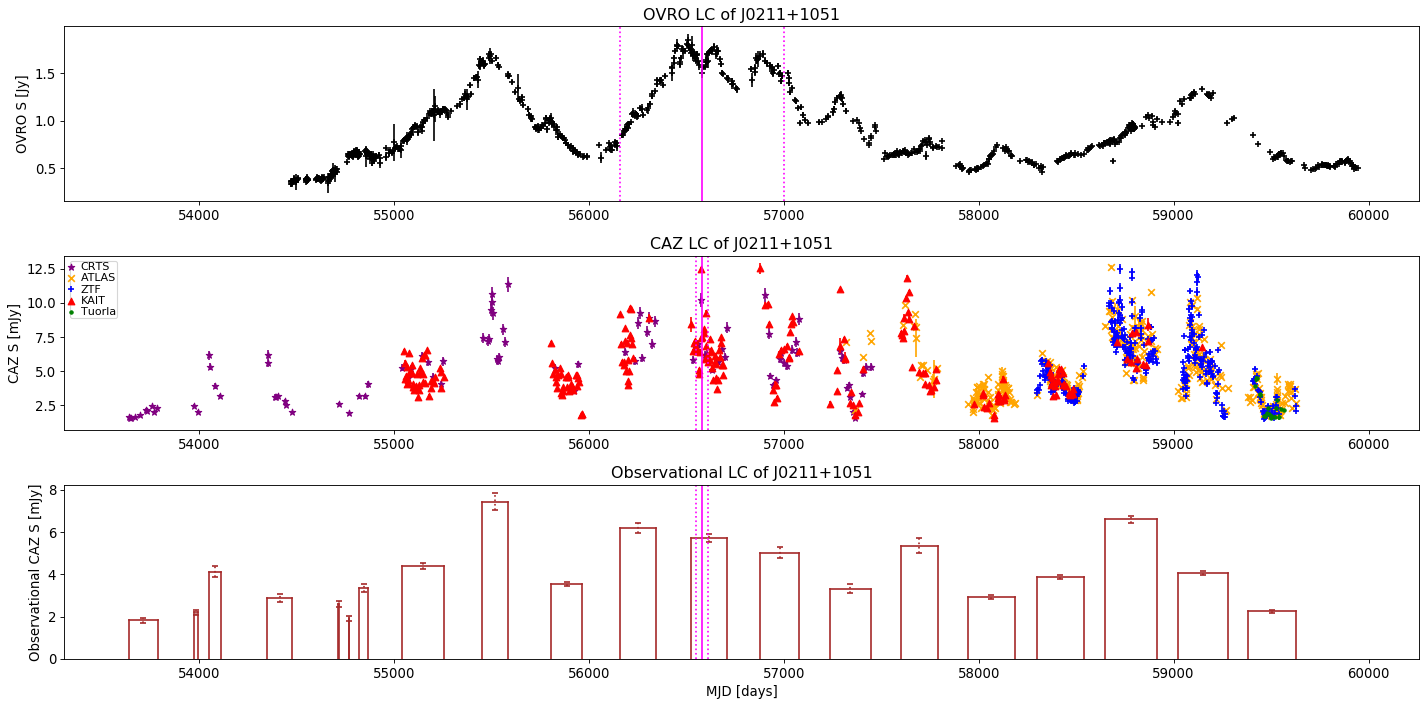}
\caption{The OVRO (radio), CAZ (optical), and observational CAZ light curves of the blazar J0211+1051 (alias: 5BZB~J0211+1051, 4FGL~J0211.2+1051) are given from top to bottom, respectively. The horizontal axis gives time in Modified Julian Date (MJD) and the vertical axis gives flux density (given in Jy in case of radio and mJy in case of optical). The vertical lines show the arrival time of a spatially associated neutrino, IC131014A. The dotted vertical lines show the AI time window centered around the neutrino arrival time. This is a spatially accurate neutrino event with  $\Omega_{\mathrm{enlarged}}$=4.48 deg$^2$ and $\mathcal{S}$=0.665. For this event, the blazar was flaring in the radio (AI$_\mathrm{R}$=1.87 \& wAI$_\mathrm{R}$=1.25) and optical (AI$_\mathrm{O}$=1.47 \& wAI$_\mathrm{O}$=0.98) bands simultaneously. This association, J0211+1051--IC131014A, only occurs when considering the enlarged error region scenario because the blazar is outside of the IceCube reported 90\% confidence level error region by 0.16\degree.}
\label{fig:J0211+1051}
\end{figure*}

\begin{figure*}
\centering
\includegraphics[width=18cm, keepaspectratio]{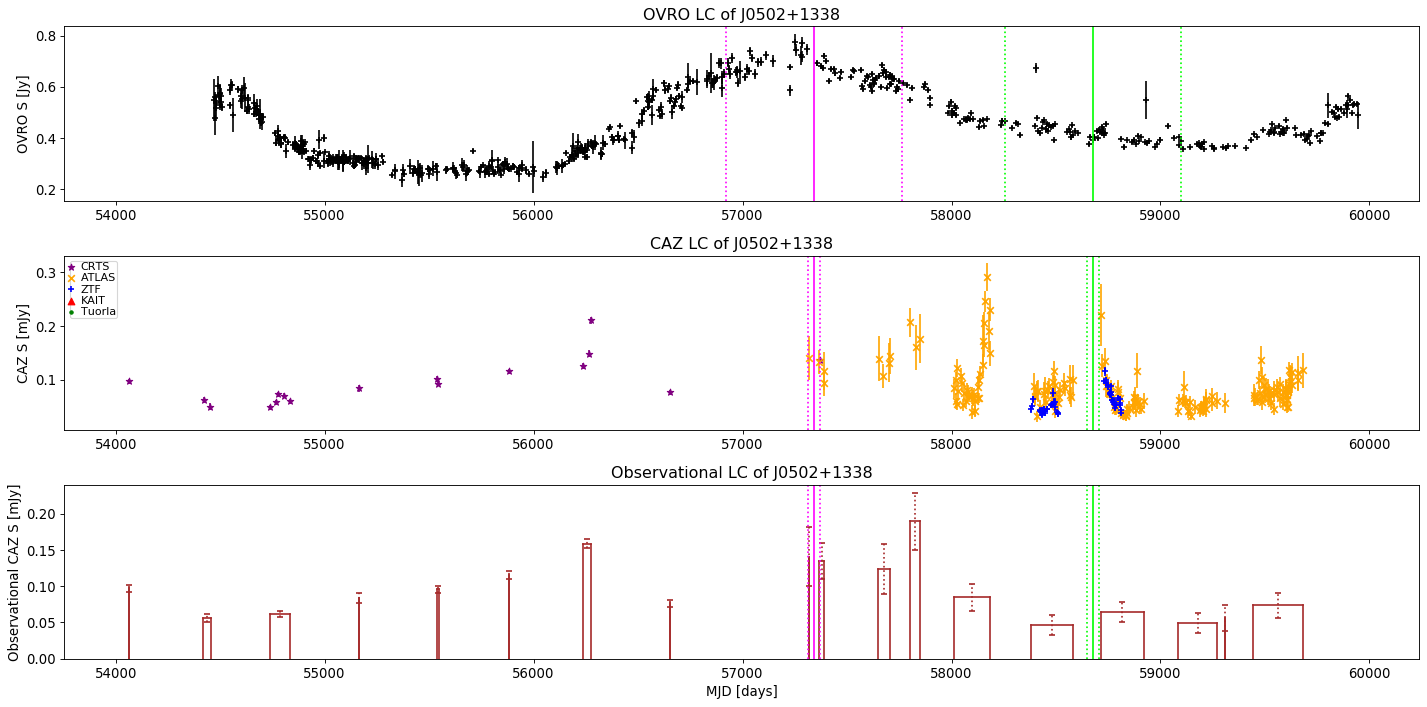}
\caption{The OVRO (radio), CAZ (optical), and observational CAZ light curves of the blazar J0502+1338 (alias: 5BZB~J0502+1338, 4FGL~J0502.5+1340) are given from top to bottom, respectively. The horizontal axis gives time in Modified Julian Date (MJD) and the vertical axis gives flux density (given in Jy in case of radio and mJy in case of optical). The vertical lines show the arrival time of spatially associated neutrinos, IC151114A (pink) and IC190712A (green). The dotted vertical lines show the AI time window centered around the neutrino arrival time. The former event is a spatially accurate one with $\Omega_{\mathrm{enlarged}}$=6.45 deg$^2$ and $\mathcal{S}$=0.957. For this event, the blazar was flaring in the radio (AI$_\mathrm{R}$=1.61 \& wAI$_\mathrm{R}$=1.54) and optical (AI$_\mathrm{O}$=1.52 \& wAI$_\mathrm{O}$=1.46) bands simultaneously. This association, J0502+1338--IC151114A, only occurs when considering the enlarged error region scenario because the blazar is outside of the IceCube reported 90\% confidence level error region by 0.34\degree.}
\label{fig:J0502+1338}
\end{figure*}

\begin{figure*}
\centering
\includegraphics[width=18cm, keepaspectratio]{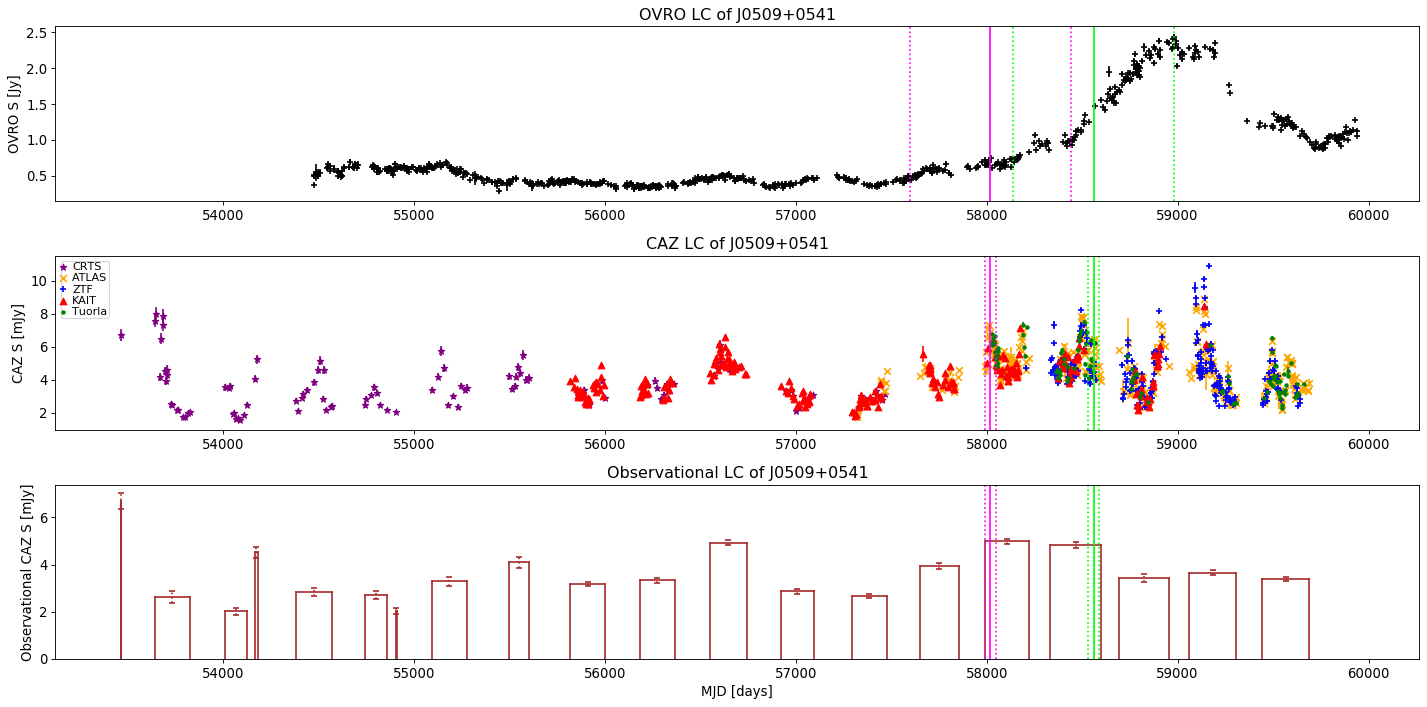}
\caption{The OVRO (radio), CAZ (optical), and observational CAZ light curves of the famous blazar J0509+0541 (alias: 5BZB~J0509+0541, 4FGL~J0509.4+0542, TXS~0506+056) are given from top to bottom, respectively. The horizontal axis gives time in Modified Julian Date (MJD) and the vertical axis gives flux density (given in Jy in case of radio and mJy in case of optical). The vertical lines show the arrival time of spatially associated neutrinos, IC170922A (pink) and IC190317A (green). The dotted vertical lines show the AI time window centered around the neutrino arrival time. The former event is a spatially accurate one with $\Omega_{\mathrm{enlarged}}$=4.93 deg$^2$ and $\mathcal{S}$=0.631 which lead to the most famous individually associated blazar-neutrino spatio-temporal connection to-date \citep{icecube_collab2018}; in our study this is also a very prominent flaring association but only in the optical band (see \S \ref{sec:results_ind_assoc}). The latter neutrino event has notably less accurate spatial resolution with $\Omega_{\mathrm{enlarged}}$=77.96 deg$^2$ and $\mathcal{S}$=0.26, which leads to the weakest (after weighting) of the four radio+optical flaring associations. Note that both of these associations occur in both the published and enlarged error region scenarios.}
\label{fig:J0509+0541}
\end{figure*}

\begin{figure*}
\centering
\includegraphics[width=18cm, keepaspectratio]{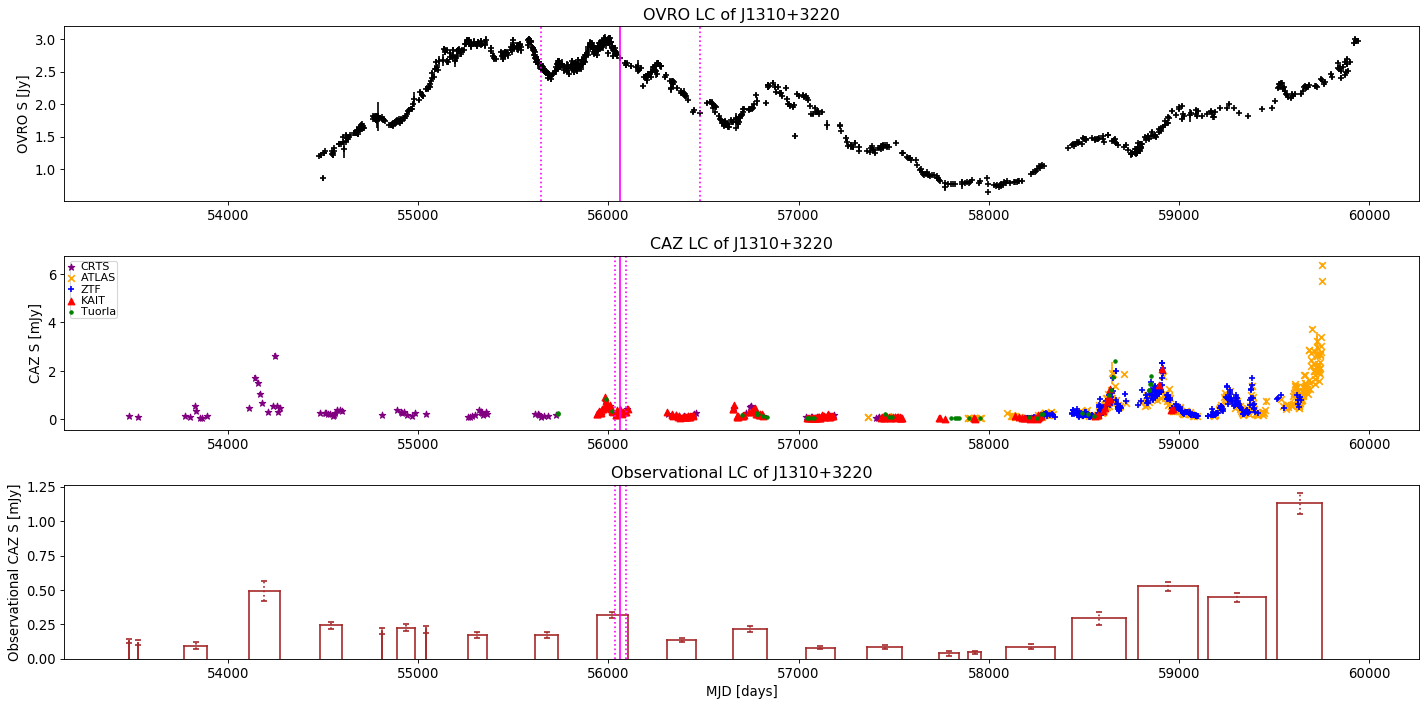}
\caption{The OVRO (radio), CAZ (optical), and observational CAZ light curves of the blazar J1310+3220 (alias: 5BZU~J1310+3220, 4FGL~J1310.5+3221, B2~1308+32.6) are given from top to bottom, respectively. The horizontal axis gives time in Modified Julian Date (MJD) and the vertical axis gives flux density (given in Jy in case of radio and mJy in case of optical). The vertical lines show the arrival time of a spatially associated neutrino, IC120515A. The dotted vertical lines show the AI time window centered around the neutrino arrival time. This is a somewhat spatially accurate neutrino event with  $\Omega_{\mathrm{enlarged}}$=8.37 deg$^2$ and $\mathcal{S}$=0.613. For this event, the blazar was flaring in the radio (AI$_\mathrm{R}$=1.42 \& wAI$_\mathrm{R}$=0.87) and optical (AI$_\mathrm{O}$=1.29 \& wAI$_\mathrm{O}$=0.79) bands simultaneously. This association, J1310+3220--IC120515A, only occurs when considering the enlarged error region scenario because the blazar is just outside of the IceCube reported 90\% confidence level error region by 0.01\degree.}
\label{fig:J1310+3220}
\end{figure*}

\subsubsection{Comments on individual flaring associations} \label{sec:results_ind_assoc}
In this subsection we discuss a few individual confidently (99.99\% confidence level) flaring neutrino-associated blazars of particular interest (see Table \ref{tab:flr_assoc} for a complete list of the most confidently flaring associations and Appendix \ref{appendix:all_flr_LCs} for all their blazar light curves).
Using the enlarged error region scenario, there are four blazars that exhibit flaring behavior simultaneously in the radio and optical bands at the time of the neutrino arrival:  
J0211+1051, J0502+1338, J0509+0541 (aka TXS~0506+056) and J1310+3220. Their light curves are shown in Figures \ref{fig:J0211+1051}, \ref{fig:J0502+1338}, \ref{fig:J0509+0541}, and \ref{fig:J1310+3220}, respectively.

\begin{itemize}
\item {\bf J0211+1051:}
Although both the radio and optical light curves show many flares, the neutrino (IC131014A) arrived around the time of the most prominent radio flare with an AI value of 1.85 (in H21, the AI value was reported to be 1.87).
This is the second most prominent strongly flaring blazar-neutrino association in the radio band.
In the optical band the flares have rather short timescales and it seems that the neutrino arrived during a fast optical flare.
As the overall optical flux of that observational block was high, this flaring association ends up as the third most prominent flaring association in the optical band.
This source is also detected in $\gamma$-ray band by {\it Fermi}-LAT and according to \citet{righi19} it showed modest activity (but no a major flare) around the time of the neutrino arrival.

\item {\bf J0502+1338:}
This blazar is associated with the neutrino IC151114A and has the highest weight of all neutrinos associated with flaring blazars.
This neutrino arrived during a very prominent radio flare, but the optical light curve in that observational block consists of only a few data points, which nevertheless exhibits higher than average flux density level.
The blazar also has a second neutrino association (IC190712A), but with much poorer spatial accuracy ($\Omega_{\mathrm{enlarged}}$=76.71 deg$^2$) and lower signalness ($\mathcal{S}$=0.304).
In the radio band the source was not flaring when the second neutrino arrived (AI$_\mathrm{R}$=0.97), while in the optical band AI information is not available as the neutrino arrived during an observational gap.

\item {\bf J0509+0541 (aka TXS~0506+056):}
This is the only blazar that has a simultaneous radio+optical flaring association (with the neutrino IC190317A) in the published error region scenario. 
However, it should be noted that this is a rather poorly reconstructed neutrino event ($\Omega_{\mathrm{enlarged}}$=77.96 deg$^2$ and $\mathcal{S}$=0.26) with a small weight (0.035) and has an association also with four other blazars.
This blazar has another neutrino (IC170922A) association which is the most significant individual blazar-neutrino association up to date (\citealt{icecube_collab2018}).
Although this neutrino event arrived during a period of heightened optical activity (AI$_\mathrm{O}$=1.42) resulting in a very prominent flaring association, in the radio band it arrived just at the start of a major radio flare.
As already discussed in H21, the AI value suggests the source to be non-flaring (AI$_\mathrm{R}$<1) as the majority of the flare falls outside of the 2.3 year time window.

\item {\bf J1310+3220:}
This blazar is associated with the neutrino IC120515A, whose weight is in the top five of both the radio-only and optical-only flaring associations shown in Table \ref{tab:flr_assoc}.
In the radio band, the flare is very prominent.
In the optical band, there is also a flare in the same observational block that the neutrino arrived in; however, there are several more prominent flares at the end of the light curve.
\end{itemize}

In addition to the blazar-neutrino associations which exhibit simultaneous radio and optical flares at the time of the neutrino arrival, there are also a number of associations with flaring activity only in one of the bands whose neutrino has a high weight:

\begin{itemize}
\item {\bf J1504+1029 (aka PKS 1502+106):}
The associated neutrino (IC190730A) of this blazar arrived at the time of a long-lasting major radio flare as already discussed in \cite{kiehlman19} and H21.
In the optical band the source was in a low state (AI$_\mathrm{O}$=0.94).
The source was also in a low state in the $\gamma$-ray band (e.g. \citealt{Franckowiak20, Rodrigues21}).

\item {\bf J1623+3909:}
This blazar is associated with the neutrino IC190201A, which was a time of high radio activity for the blazar (AI$_\mathrm{R}$=1.31).
AI$_\mathrm{O}$ of 1.09 is indicative of a higher than average flux density at the time of the neutrino arrival.

\item {\bf J2030-0503:}
This blazar is associated with the neutrino IC180417A and its AI$_\mathrm{O}$ is 2.33 before weighting (the highest in our sample).
The neutrino arrived just before the start of the ZTF observations and the first observational block contained a major optical flare while in the other observational blocks the source was in a low state. 
\end{itemize}

We note that while the four simultaneous radio+optical flaring associations give the strongest correlation significances between the CGRaBS blazars and IceCat1+ neutrinos, there are many more radio-only and optical-only associations with comparable AI values as shown in Table \ref{tab:flr_assoc}.
As such we do not claim that the simultaneity of blazar light curve activity in the radio and optical bands is a general signature of blazar-neutrino correlation.

\subsubsection{Comparing the results of the published and the enlarged scenarios} \label{sec:results_minimal_vs_maximal}
According to the p-values given in Table \ref{tab:final_pvals}, the CGRaBS blazars are not correlated with IceCat1+ neutrinos in the published error region scenario.
However, after increasing the neutrino error regions by adding 1\degree\, in quadrature to each of the four error bars, five individual correlation significances increase up to 2$\sigma$, one of which even reaches the 3$\sigma$ level.
This is due to the existence of several blazars just outside of the published error regions of some of the most accurately reconstructed events that were strongly flaring in the radio and/or optical bands when the neutrino arrived.

Column (6) in Table \ref{tab:flr_assoc} shows the associations, marked with "*", that only occur after increasing the IceCube error regions.
Using the IceCube error regions without accounting for any systematic errors, only two and one of the most prominent flaring associations in the radio and optical bands would remain, respectively.
Of the four simultaneous radio+optical flaring associations driving our strongest individual correlation significance (>3$\sigma$, see Table \ref{tab:final_pvals}), only one occurs within the published error regions, which is also the weakest of the four associations due to its neutrino event being rather unreliable ($\Omega_\mathrm{published}$=74.6 deg$^2$ and $\mathcal{S}$=0.26). On the other hand, the three remaining associations (arising from some of the most reliable neutrino events in the sample) only occur in the enlarged scenario --- they miss their respective published error region by 0.34\degree, 0.08\degree, and 0.01\degree. In the case of AI$_{0.01\%}$ as TS parameter (resulting in the most significant individual p-values of this study), as evident from columns (6) and (7) of Table \ref{tab:flr_assoc}, all of the associations that only exist because of enlarging the error regions, fall outside of the published error regions by much less than 1\degree.

Two recent studies (\citealt{abbasi+2023_plavin_test, bellenghi+2023_5bzcat_and_rfc_expansion}) have indicated that the $\sim$3$\sigma$ spatial blazar-neutrino correlations obtained using the enlarged error regions could be the result of a random fluctuation in the blazar spatial distribution. Our study is significantly different as we not only look at spatial associations (the case of $\overline{S}$ as TS parameter), but also incorporate temporal and multi-band elements into the statistics. Still, we independently find a 2.17$\sigma$ (post-trial) blazar-neutrino correlation in the enlarged scenario. As discussed, \cite{abbasi+2023_plavin_test} stated that the IceCube collaboration still cannot exclude the possibility of unknown systematic uncertainties in the neutrino localization. Assuming that the reported error regions are larger by 1\degree\, in quadrature (the estimated upper limit) due to these uncertainties, then our results hint at a spatio-temporal connection between the multiwavelength emission of CGRaBS blazars and the IceCube HE neutrino events.

\section{Summary \& Conclusions} \label{sec:conclusion}
In this paper, we studied a spatio-temporal correlation between the CGRaBS blazars and the IceCube neutrino events using the radio and optical bands.
We used the most up-to-date neutrino list released by the IceCube collaboration in 2023 with a total of 283 HE ($\gtrsim$100 TeV) neutrino events, see \S \ref{sec:data_neut}.
The CGRaBS sample consists of 1157 blazars with OVRO 15 GHz radio light curves from 2008 until the end of 2022.
For most (1061) of these blazars, we managed to combine data from all-sky surveys and dedicated monitoring programs in the optical band to create decade-long light curves, see \S \ref{sec:data_blz}.

In \cite{hovatta+2021}, H21, our main conclusion was that observations of large-amplitude radio flares in blazars spatially-associated with neutrinos at the time of their arrival are unlikely to occur by random chance ($\sim$2$\sigma$).
In this work, we find that adding three more years of neutrino data and two more years of radio data to the 12 years used in H21 does not improve the correlation significances. 
Upon upgrading the spatio-temporal method to take neutrino signalness and error region sizes into account, we find that our radio results remain consistent with H21.
Our optical results are also similar.
The highest individual significance (>3$\sigma$) of the spatio-temporal correlation between the CGRaBS blazars and IceCube neutrinos occurs when we consider the number of spatially-coincided blazars that are confidently flaring in both the radio and optical bands simultaneously at the time of the neutrino arrival.
In \S \ref{sec:trial_correction}, we discussed the post-trial correction of obtaining five >2$\sigma$ individual p-values (one of which is the aforementioned >3$\sigma$ p-value) by chance via calculating the harmonic mean p-value, where a 2.17$\sigma$ spatio-temporal correlation between the multiwavelength emission of CGRaBS blazars and IceCube HE neutrinos was established. However, we note that this 2.17$\sigma$ correlation only occurs when we add the estimated upper limit of the IceCube systematic errors (1\degree) in quadrature to all the error bars. When the IceCube error regions are taken as their published values, we find no significant correlation.
The observed improvement in significances when considering larger neutrino error regions, as discussed in \S \ref{sec:results_minimal_vs_maximal}, is in line with the results of \cite{abbasi+2023_plavin_test}.

In addition to the unclear systematic errors of the neutrino events, another major challenge for population studies is the inclusion of neutrino events with very large error regions.
In H21 we used thresholds on the energy and the error region size, while in this work we used a dedicated weighting scheme.
The main motivation for this was to refrain from making arbitrary choices in the data selection, to preserve the signal from the events close to the thresholds, and to boost the resilience of the statistics against slight event parameter changes, while simultaneously reducing the noise contribution from the most poorly reconstructed events.
However, the noise suppression arising from using thresholds is more efficient and, as a result, we suspect that the optimal way of performing the spatio-temporal analysis is to employ both the thresholds and the weighting scheme simultaneously (i.e., by cutting out the highest error region events that have tens of associated blazars), while weighting the rest of the neutrino events according to their signalness and the error region size.
Nevertheless, we conservatively chose not to pursue such aggressive selection and weighting schemes in this work, because those would push our statistics to a more individualistic direction where a few of the most well-defined neutrino events would drive our statistics.
Moreover, determining where to place the threshold would require careful simulations and is beyond the scope of this paper.

Furthermore, we note that while the OVRO-monitored CGRaBS blazar sample used in this study provides the most complete sample of blazar radio light curves with a high temporal accuracy and cadence, it is not as complete as VLBI-based catalogs in terms of flux density, and it suffers from a blazar type bias due to its lack of HSP blazars.
Such completeness issues of the CGRaBS sample combined with the >2$\sigma$ individual correlations seen in the optical band motivated us to investigate the blazar-neutrino connection with a much larger (>4$\times$) sample of blazars in the optical band using TS parameters that probe faster than observational temporal variations (Kouch et al., in prep.).

The key takeaway points of this spatio-temporal correlation study between CGRaBS blazars and HE IceCube neutrino events are summarized below:
\begin{itemize}
    \item We developed a novel weighting scheme that allows for the utilization of the entire neutrino dataset without the need for arbitrary thresholds. 
    \item We extended the spatio-temporal analysis to the optical band and found that its results are similar to those of the radio band (i.e., spatially-associated neutrinos seem to preferentially arrive at long-term periods of higher-than-average electromagnetic flux density).
    \item We found that the most significant correlations occur when considering long-term radio and optical flares simultaneously. 
    \item Crucially, all of our significant results (>2$\sigma$) only occurred when we enlarged the published neutrino error regions by 1\degree\, in quadrature.
\end{itemize}

Notably, how the IceCube positional error regions are taken into account makes a critical difference in the final results of our spatio-temporal correlation study. If the published error regions are accurate, we see no connection between the multiwavelength electromagnetic emission of blazars and HE neutrino emission. However, if the error regions are underestimated due to unknown systematic errors (using their estimated upper limit of 1\degree, added in quadrature), then our post-trial p-value indicates a spatio-temporal, multiwavelength blazar-neutrino correlation with a significance of at least 2.17$\sigma$. Interestingly, in the case of our most significant individual p-values (when AI$_{0.01\%}$ is used as TS parameter) as seen from columns (6) and (7) of Table \ref{tab:flr_assoc}, all spatio-temporal associations that only occur in the enlarged scenario are much closer to the published error regions than 1\degree. In any case, the low association numbers suggest that not all neutrinos are associated with the most strongly flaring CGRaBS blazars, which is in line with our findings in H21.

Finally, we emphasize that to improve the prospect of definitively solving the blazar-neutrino correlation mystery, it is crucial to have neutrino events with better signalness and directional resolution.
To this end, next generation neutrino observatories such as IceCube-Gen2 \citep{aartsen2014_icecube_gen2}, Baikal-GVD \citep{shoibonov+2012_bikal_gvd_upgrade}, KM3NeT \citep{k3mnet_collab2016_next_gen}, RNO-G \citep{aguilar+2021_greenland_neut_obs}, and GRAND \citep{fang+2017_china_neut_obs} will play vital roles. \\

\begin{acknowledgements}
      P.K. was supported by Academy of Finland projects 346071 and 345899.
      E.L. was supported by Academy of Finland projects 317636, 320045, and 346071.
      T.H. was supported by Academy of Finland projects 317383, 320085, 322535, and 345899.
      J.J. was supported by Academy of Finland projects 320085 and 345899. 
      K.K. acknowledges support from the European Research Council (ERC) under the European Union’s Horizon 2020 research and innovation programme (grant agreement No. 101002352).
      S.K. acknowledges support from the European Research Council (ERC) under the European Unions Horizon 2020 research and innovation programme under grant agreement No. 771282.
      R.R. is supported by ANID BASAL grant FB210003.
      W.M. gratefully acknowledges support by the ANID BASAL project FB210003 and FONDECYT 11190853. \\
      
      The OVRO 40 m program was supported by NASA grants NNG06GG1G, NNX08AW31G, NNX11A043G, and NNX13AQ89G from 2006 to 2016 and NSF grants AST-0808050, and AST-1109911 from 2008 to 2014, along with private funding from Caltech and
      the MPIfR. \\
      
      Based on observations obtained with the Samuel Oschin Telescope 48-inch and the 60-inch Telescope at the Palomar Observatory as part of the Zwicky Transient Facility project. ZTF is supported by the National Science Foundation under Grant No. AST-2034437 and a collaboration including Caltech, IPAC, the Weizmann Institute for Science, the Oskar Klein Center at Stockholm University, the University of Maryland, Deutsches Elektronen-Synchrotron and Humboldt University, the TANGO Consortium of Taiwan, the University of Wisconsin at Milwaukee, Trinity College Dublin, Lawrence Livermore National Laboratories, and IN2P3, France. Operations are conducted by COO, IPAC, and UW. The ZTF forced-photometry service was funded under the Heising-Simons Foundation grant \#12540303 (PI: M.J.Graham). \\
      
      This work has made use of data from the Asteroid Terrestrial-impact Last Alert System (ATLAS) project. The Asteroid Terrestrial-impact Last Alert System (ATLAS) project is primarily funded to search for near earth asteroids through NASA grants NN12AR55G, 80NSSC18K0284, and 80NSSC18K1575; byproducts of the NEO search include images and catalogs from the survey area. This work was partially funded by Kepler/K2 grant J1944/80NSSC19K0112 and HST GO-15889, and STFC grants ST/T000198/1 and ST/S006109/1. The ATLAS science products have been made possible through the contributions of the University of Hawaii Institute for Astronomy, the Queen’s University Belfast, the Space Telescope Science Institute, the South African Astronomical Observatory, and The Millennium Institute of Astrophysics (MAS), Chile. \\

      This work has made use of data from the Joan Oró Telescope (TJO) of the Montsec Observatory (OdM), which is owned by the Catalan Government and operated by the Institute for Space Studies of Catalonia (IEEC).\\

      We additionally thank Konstancja Satalecka and Simone Garrappa for their insightful comments when handling the IceCube neutrino events.
\end{acknowledgements}

\bibliographystyle{aa} 
\bibliography{ref.bib} 

\begin{appendix}
\section{The reduction and combination process of the CAZ optical light curves} \label{appendix:caz_combination}
As briefly mentioned in \S \ref{sec:data_blz}, the CAZ optical light curves are made by combining CRTS, ATLAS, and ZTF (as well as whenever possible KAIT and/or Tuorla) light curves. Thus the backbone of the CAZ light curves is formed by three all-sky surveys which use forced-photometry on a specified coordinate in the sky to perform automatic reduction. While such all-sky surveys have the advantage (as compared to dedicated blazar monitoring programs) of allowing for the extraction of thousands of blazar light curves with relatively high cadence, they suffer from poor angular resolutions (at best on the order of several arcsec). For this reason we decided to use the radio coordinates of the CGRaBS blazars to the closest 0.001\degree\, (or 3.6") when obtaining their optical light curves.

In the case of CRTS and ZTF the extracted light curves are in magnitudes ($m$). Following \cite{hovatta+2014_crts_stuff}, we performed a small empirical shift on the CRTS V-band magnitudes (dimmed by 0.145) and removed noise-like intra-night variability before converting them to flux densities ($S$) using $S=S_0\cdot10^{-0.4m}$ with a zero-magnitude flux density ($S_0$) of 3640 Jy. The ZTF magnitudes were converted to flux densities using an $S_0$ of 3631 Jy. In flux density space, the CRTS and ZTF data points were binned on a nightly basis by taking the average Modified Julian Date (MJD), average flux density, and propagating the flux density error via $(\sum_{i=1}^{N}{\sigma_i}^2)^{0.5}/N^2$. Subsequently, assuming Gaussian error bars, data points with a lower detection significance than 3$\sigma$ were ignored from the light curves.

Next, when the CRTS light curve is long enough (at least 3 data points) to justify its presence in the final light curve, we converted the ZTF light curves into the CRTS V-band using the zr and zg filter information via three different rearrangements of the formula $m_\mathrm{V}-m_\mathrm{r}=0.474 \, (m_\mathrm{g}-m_\mathrm{r})+0.006$ where $m$ is the magnitude of the different filters (see \citealt{tonry+2012_ztf_conversion_to_crts}). The three rearrangements to obtain the flux density in the V-band ($S_\mathrm{V}$) along with its propagated error ($\sigma_{S_\mathrm{V}}$) are given below:
\begin{enumerate}[i]
    
    \item when $S_\mathrm{r}$ and $S_\mathrm{g}$ are known quasi-simultaneously:\\
    $S_\mathrm{V}=\mathrm{C} \, {S_\mathrm{r}}^{0.526} \,{S_\mathrm{g}}^{0.474}$\\
    $\sigma_{S_\mathrm{V}}=\mathrm{C} \, (S_\mathrm{g}/S_\mathrm{r})^{0.474} \, \sqrt{(0.526 \, \sigma_{S_\mathrm{r}})^2+(0.474 \, \sigma_{S_\mathrm{g}} \, S_\mathrm{r}/S_\mathrm{g})^2}$
    
    \item when the quasi-simultaneous flux density values are unknown, but for a known $S_\mathrm{r}$ the value of $S_\mathrm{g}/S_\mathrm{r}$ is attainable from a linear fit in the ($S_\mathrm{r}$, $S_\mathrm{g}$) plot:\\
    $S_\mathrm{V}=\mathrm{C} \, S_\mathrm{r}\, (S_\mathrm{g}/S_\mathrm{r})^{0.474}$\\
    $\sigma_{S_\mathrm{V}}=\mathrm{C} \, \phi^{0.474} \sqrt{{\sigma_{S_\mathrm{r}}}^2+(0.474 \, \sigma_\phi \,  S_\mathrm{r}/\phi)^2}$
    
    \item when the quasi-simultaneous flux density values are unknown, but for a known $S_\mathrm{g}$ the value of $S_\mathrm{g}/S_\mathrm{r}$ is attainable from a linear fit in the ($S_\mathrm{r}$, $S_\mathrm{g}$) plot:\\
    $S_\mathrm{V}=\mathrm{C} \, S_\mathrm{g} \, (S_\mathrm{g}/S_\mathrm{r})^{-0.526}$\\
    $\sigma_{S_\mathrm{V}}=\mathrm{C} \, \phi^{-0.526} \sqrt{(0.526 \, \sigma_\phi \, S_\mathrm{g}/\phi)^2+{\sigma_{S_\mathrm{g}}}^2}$
    
\end{enumerate}
where $\mathrm{C} = 10^{-0.0024} \cdot 3640/3631 \approx 0.99695$, $\phi$ refers to the gradient of the linear fit in the ($S_\mathrm{r}$, $S_\mathrm{g}$) plot, $\sigma_\phi$ refers to the error of the gradient, and quasi-simultaneous means that there are both zr and zg data points within two nights. The ZTF light curve is ignored when quasi-simultaneous zr and zg data points are unavailable and a fit in the ($S_\mathrm{r}$, $S_\mathrm{g}$) plot is unattainable. In the cases where CRTS is ignored, the ZTF filter with the most number of data points is selected and no ZTF conversion into the V-band is needed.

In the case of ATLAS, the extracted data comes in the flux density space as well as the magnitude space, so no conversion is needed. Firstly, the most obvious ATLAS outliers are removed (i.e., those data points that have negative flux density values and those whose flux density value sits >10$\sigma$ off of the light curve mean). This 10$\sigma$ off-mean limit was chosen after careful manual inspection (while also taking advantage of data from other surveys, when available) to ensure that as few data points from regions of locally enhanced flux density (possibly signifying real flaring behavior) as possible were removed. Such a cleaning procedure is especially needed to ensure the reliability of the ATLAS shift onto ZTF/CRTS (see below). Afterwards the ATLAS light curve is nightly binned. Another round of cleaning removes data points which do not have at least a 3$\sigma$ detection significance (assuming Gaussian error bars).

Subsequently, the cleaned ATLAS light curve (in the filter that has the most number of data points) is shifted onto the ZTF or CRTS (if ZTF is not present) light curves by finding the average value of simultaneous $S_\mathrm{ZTF}/S_\mathrm{ATLAS}$, referred to as the correction factor. The rarity of nights with simultaneous data make the correction factor rather prone to individual extreme outliers, which is why we ensured to remove such extremities. The shift is done by multiplying the flux densities and their errors by the correction factor. If there are no simultaneous (daily) data points between ZTF and ATLAS, then other surveys such as the KAIT and Tuorla are used to shift them. Once ZTF and ATLAS (as well as possibly KAIT and/or Tuorla) are shifted onto one another, the combined light curve is shifted onto CRTS; either by looking for simultaneous data (this is rare as the possible overlap period between CRTS and any other survey is short) or by utilizing the flux density averages in the overlap period to obtain an estimated correction factor. Correction factors of up to 3 are accepted while larger values are considered problematic. In problematic cases no shifting is performed and the light curve is limited to the combined ZTF+ATLAS era or to the CRTS era depending on whichever has a greater number of data points. Finally, a manual inspection of the light curves is performed where the remaining outliers are removed and suspicious correction shifts are reversed.

We caution the reader that the flux density shifts are not always made to the same filter (as CRTS is not always present). This means that the absolute flux densities in the light curves where CRTS is present and in the light curves where CRTS is absent are not comparable. However, as our study focuses on variability, this caveat does not affect our results.

\section{Light curves of confidently flaring neutrino associated blazars} \label{appendix:all_flr_LCs}
\begin{figure*}
\centering
\includegraphics[width=8cm, keepaspectratio]{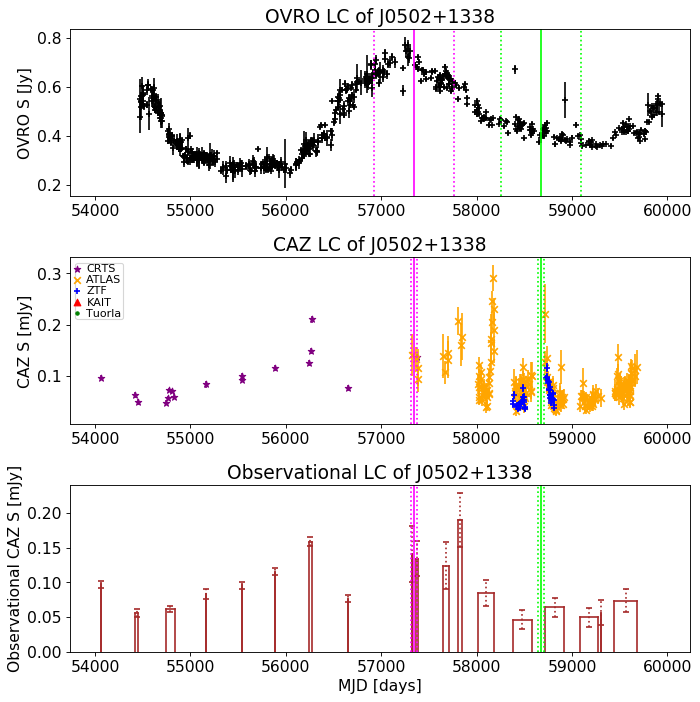}
\includegraphics[width=8cm, keepaspectratio]{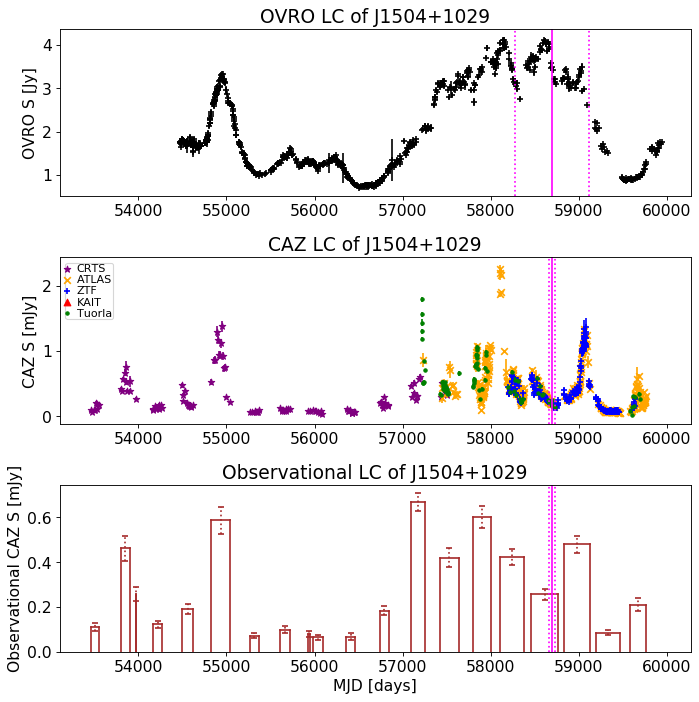}
\includegraphics[width=8cm, keepaspectratio]{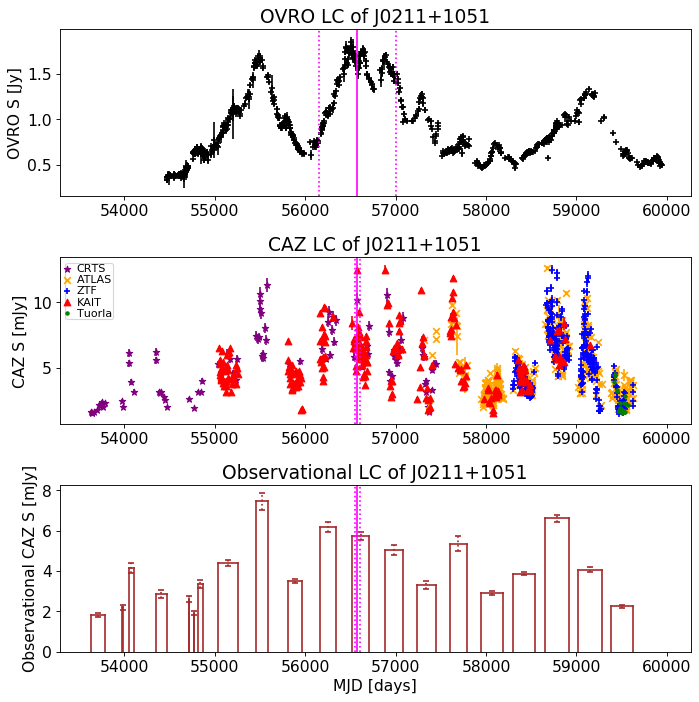}
\includegraphics[width=8cm, keepaspectratio]{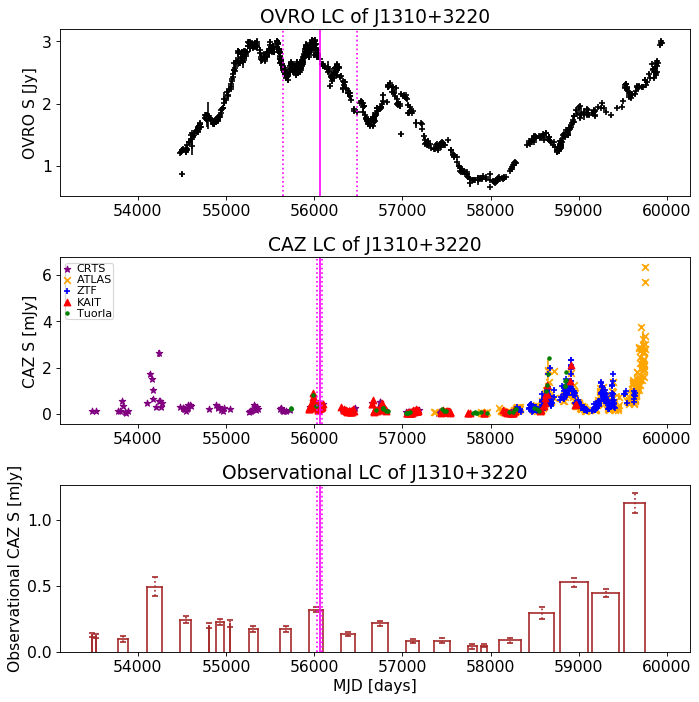}
\includegraphics[width=8cm, keepaspectratio]{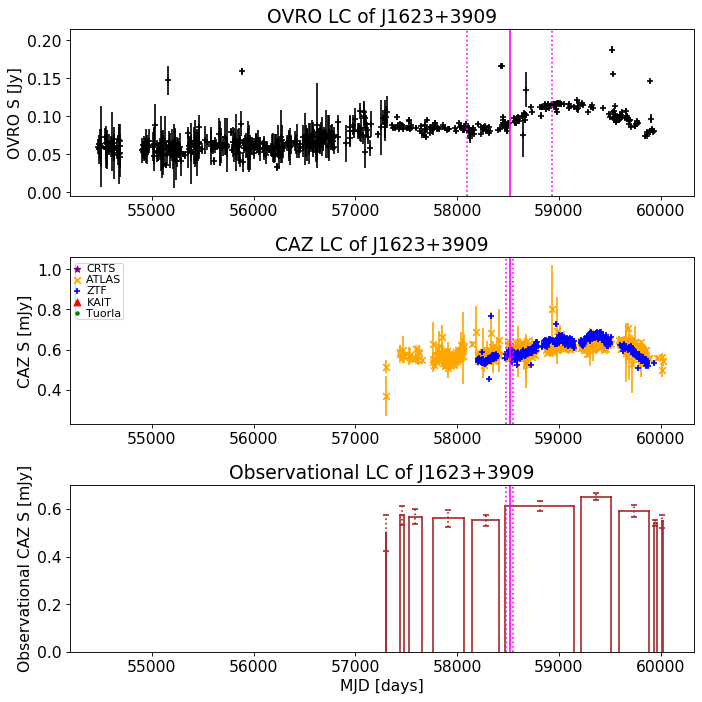}
\includegraphics[width=8cm, keepaspectratio]{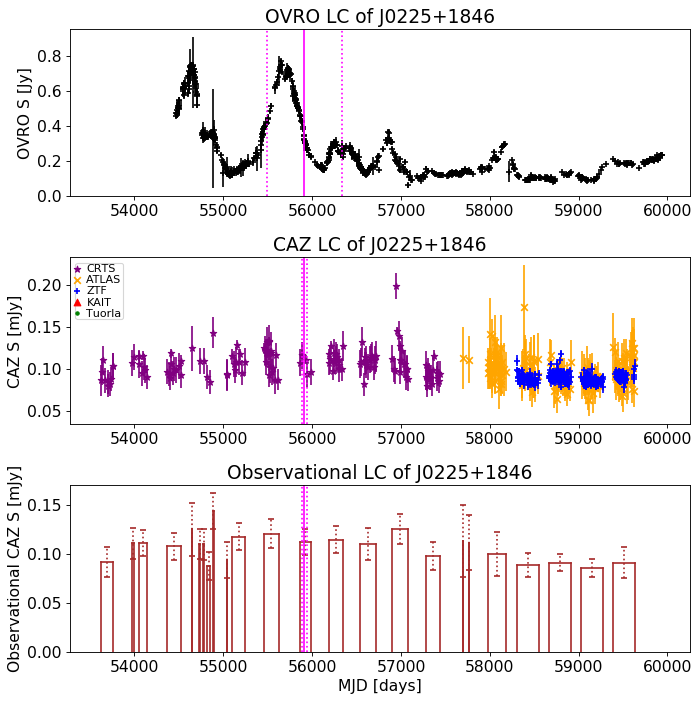}
\caption{The blazar light curves of the confidently-flaring (99.99\% confidence level) blazar-neutrino associations given in Table \ref{tab:flr_assoc}.}
\label{fig:all_flr_LCs1}
\end{figure*}

In this Appendix, in Figure \ref{fig:all_flr_LCs1}, we provide the blazar light curves of all confidently-flaring (when the associated AI value is greater than the 0.01\% false-positive rate AI threshold) blazar-neutrino associations given in Table \ref{tab:flr_assoc}.
For each blazar there are three light curves plotted: the OVRO (radio) light curve in the top panel, the CAZ (optical) light curve in the middle panel, and the observational CAZ light curve in the bottom pannel.  The horizontal axis gives the time in Modified Julian Date (MJD) while the vertical axis gives the flux density (given in Jy in case of radio and mJy in case of optical). Each solid vertical line demarcates the arrival time of a spatially-associated neutrino and the dotted vertical lines on either side of it show the AI time window centered around the neutrino arrival time.

\begin{figure*}
\centering
\includegraphics[width=8cm, keepaspectratio]{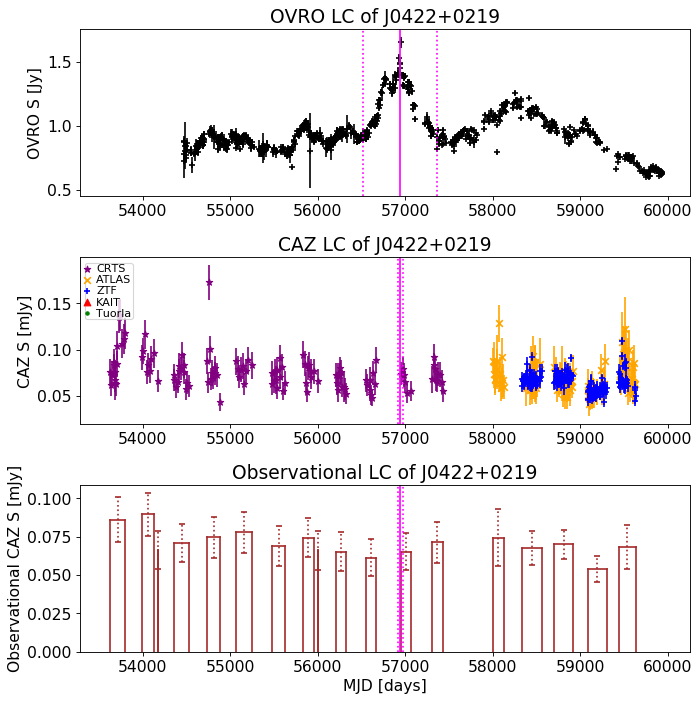}
\includegraphics[width=8cm, keepaspectratio]{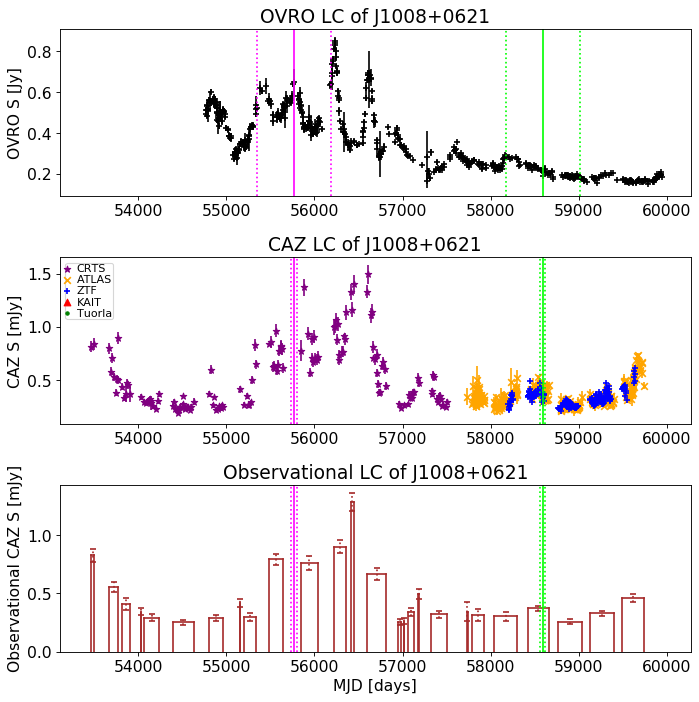}
\includegraphics[width=8cm, keepaspectratio]{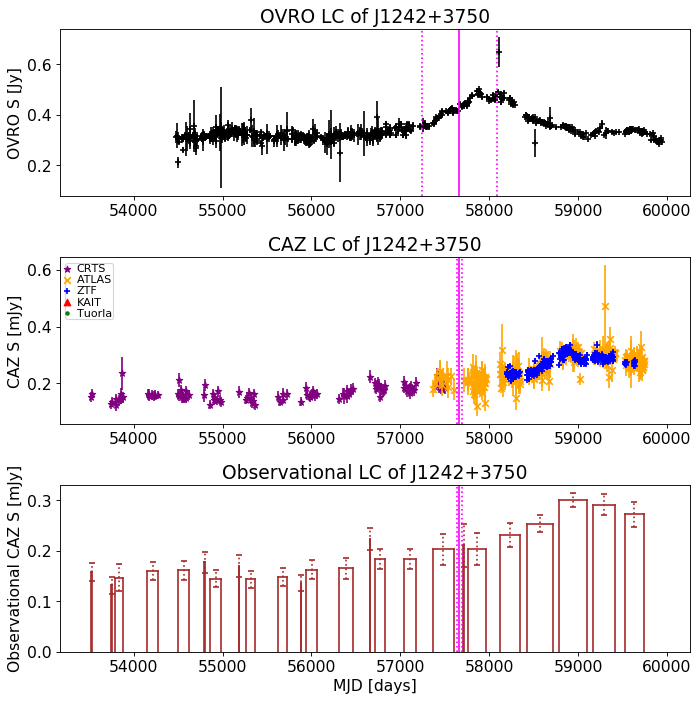}
\includegraphics[width=8cm, keepaspectratio]{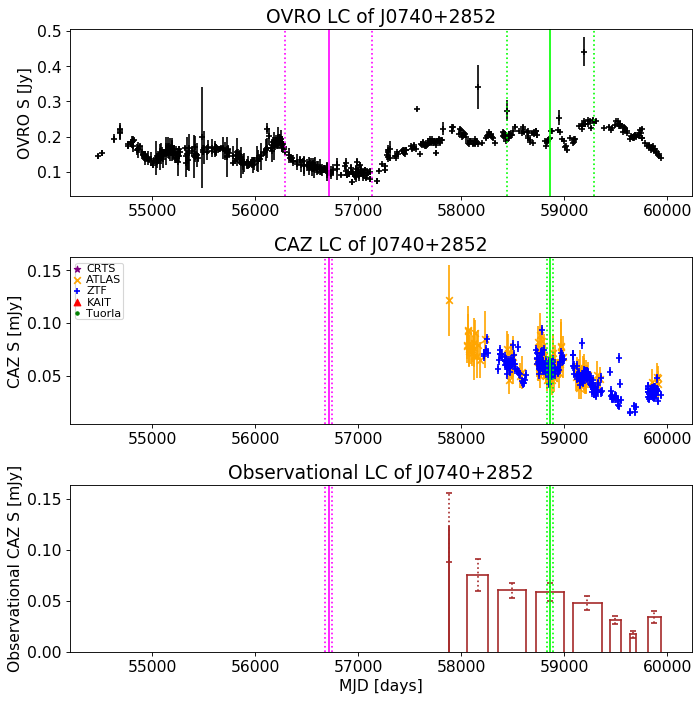}
\includegraphics[width=8cm, keepaspectratio]{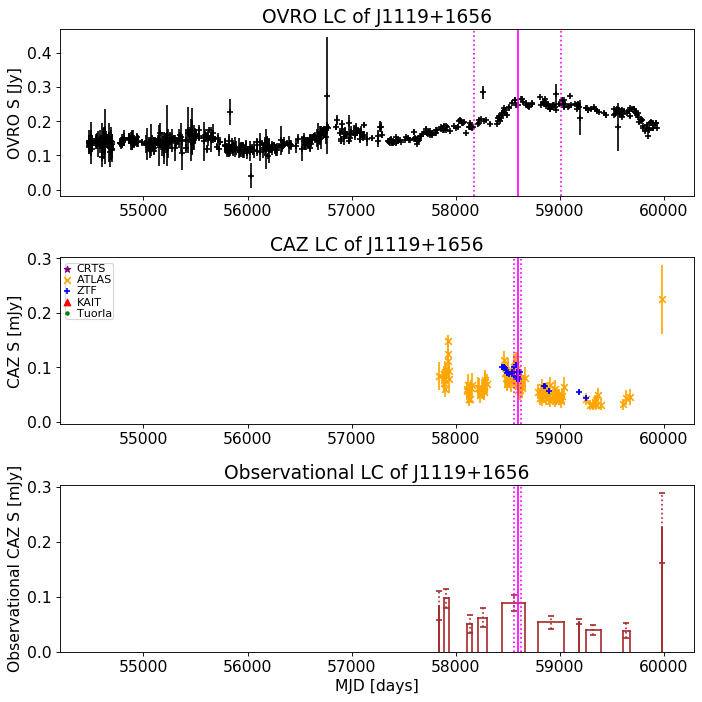}
\includegraphics[width=8cm, keepaspectratio]{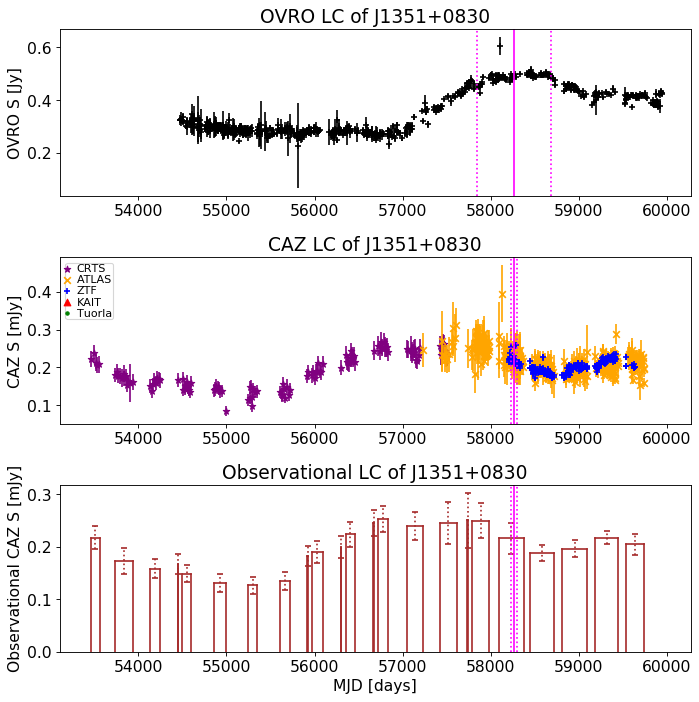}
\\
\raggedright
\textbf{Fig. B.1} continued.
\label{fig:all_flr_LCs2}
\end{figure*}

\begin{figure*}
\centering
\includegraphics[width=8cm, keepaspectratio]{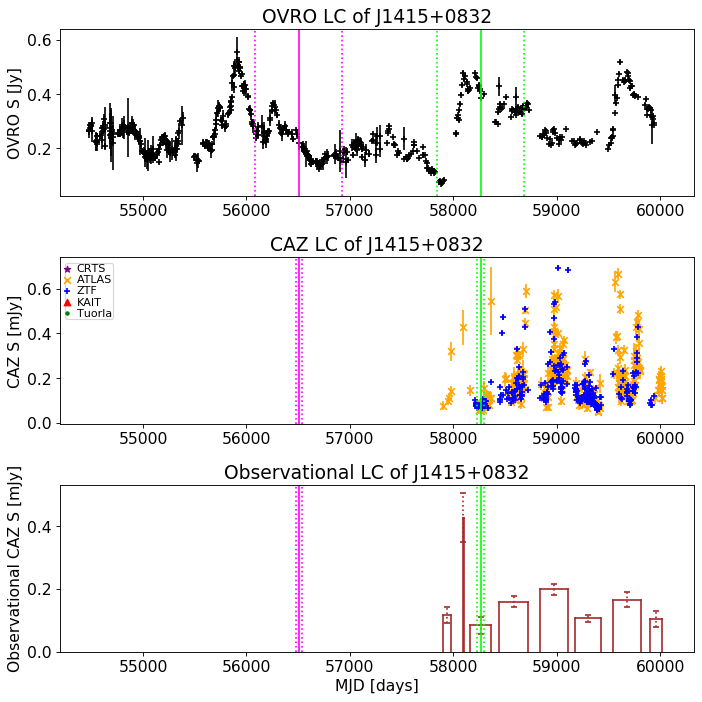}
\includegraphics[width=8cm, keepaspectratio]{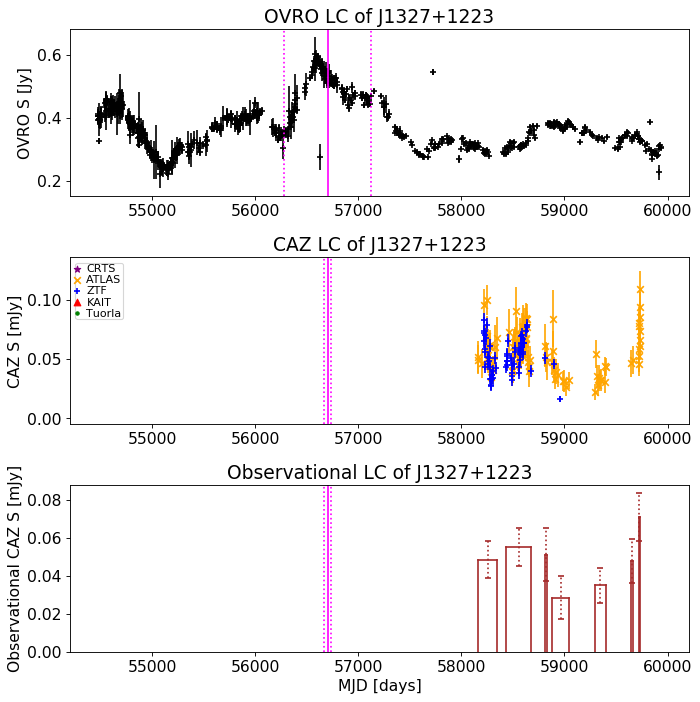}
\includegraphics[width=8cm, keepaspectratio]{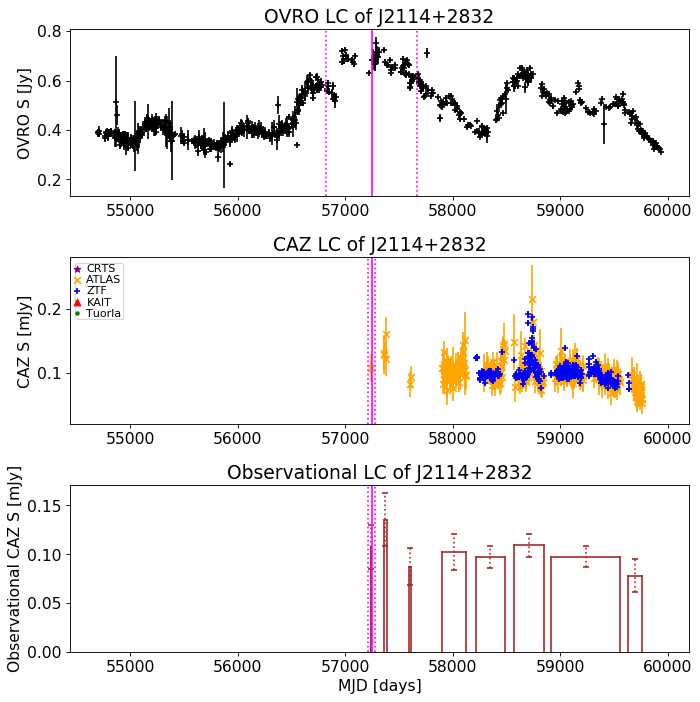}
\includegraphics[width=8cm, keepaspectratio]{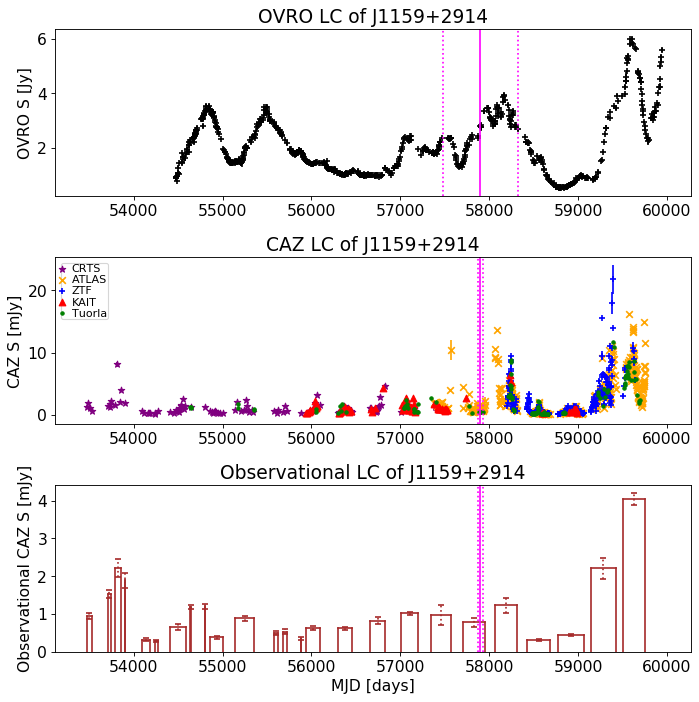}
\includegraphics[width=8cm, keepaspectratio]{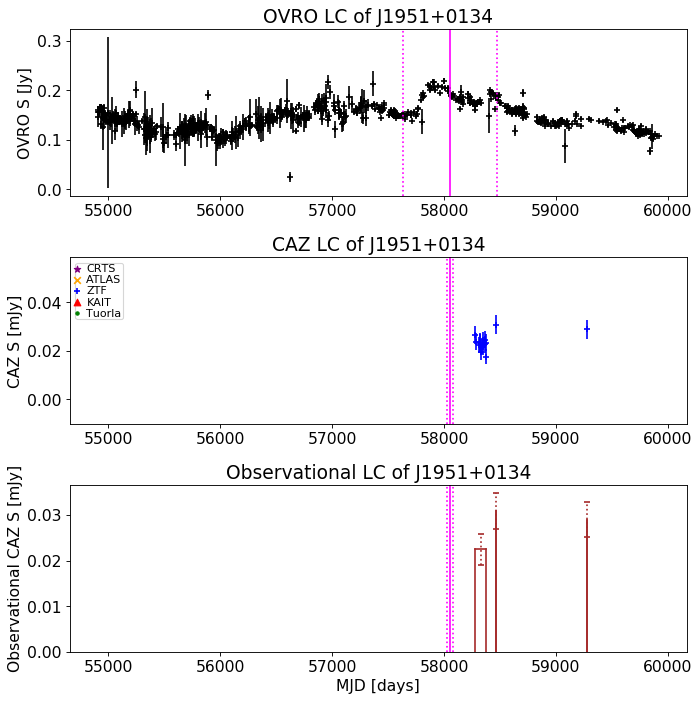}
\includegraphics[width=8cm, keepaspectratio]{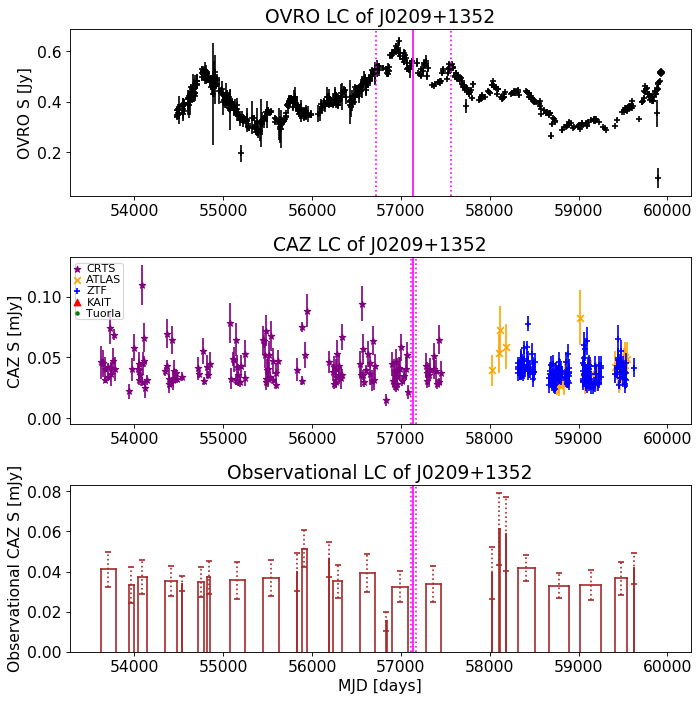}
\\
\raggedright
\textbf{Fig. B.1} continued.
\label{fig:all_flr_LCs3}
\end{figure*}

\begin{figure*}
\centering
\includegraphics[width=8cm, keepaspectratio]{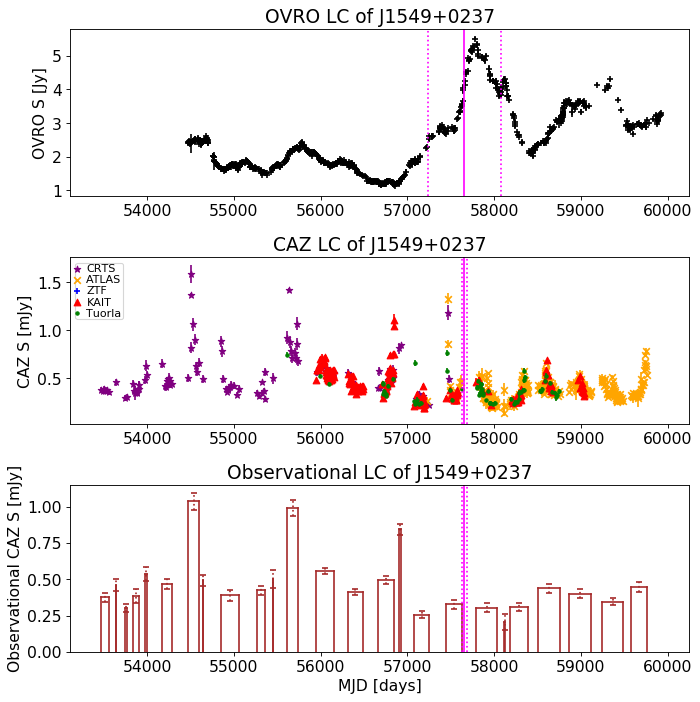}
\includegraphics[width=8cm, keepaspectratio]{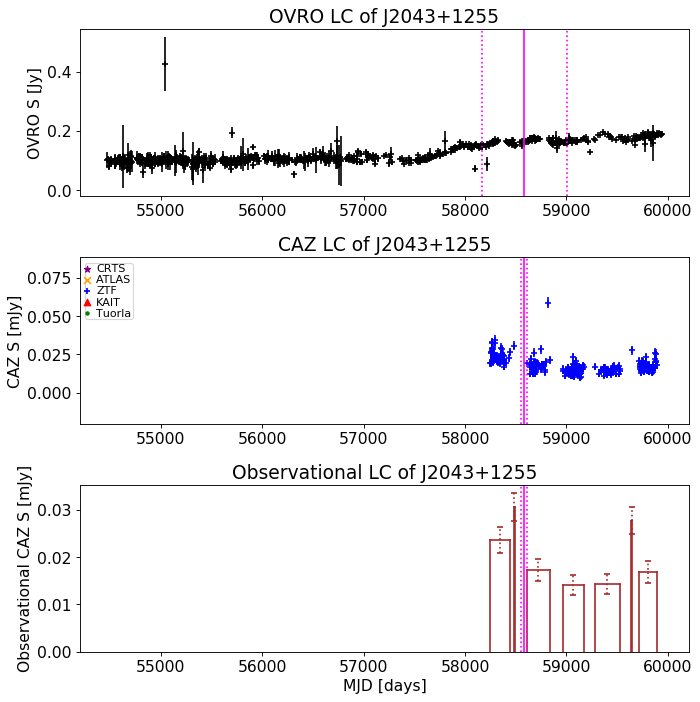}
\includegraphics[width=8cm, keepaspectratio]{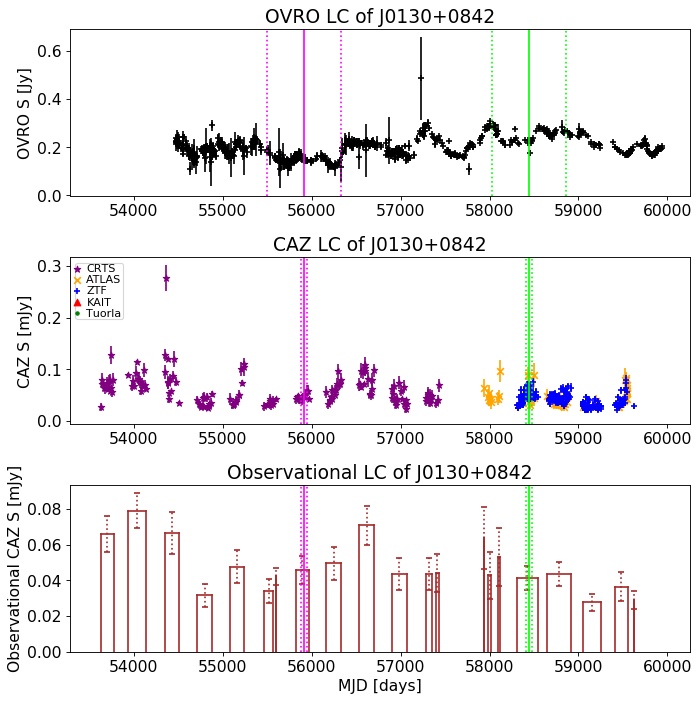}
\includegraphics[width=8cm, keepaspectratio]{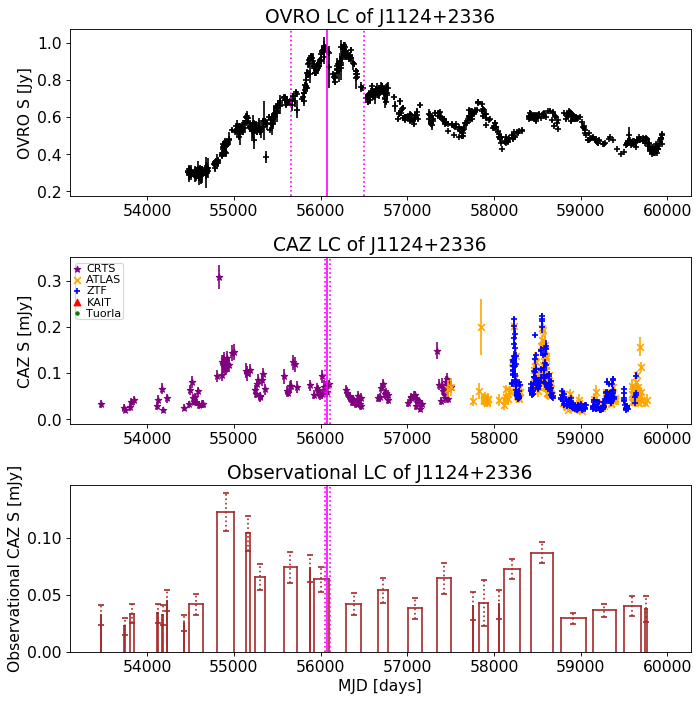}
\includegraphics[width=8cm, keepaspectratio]{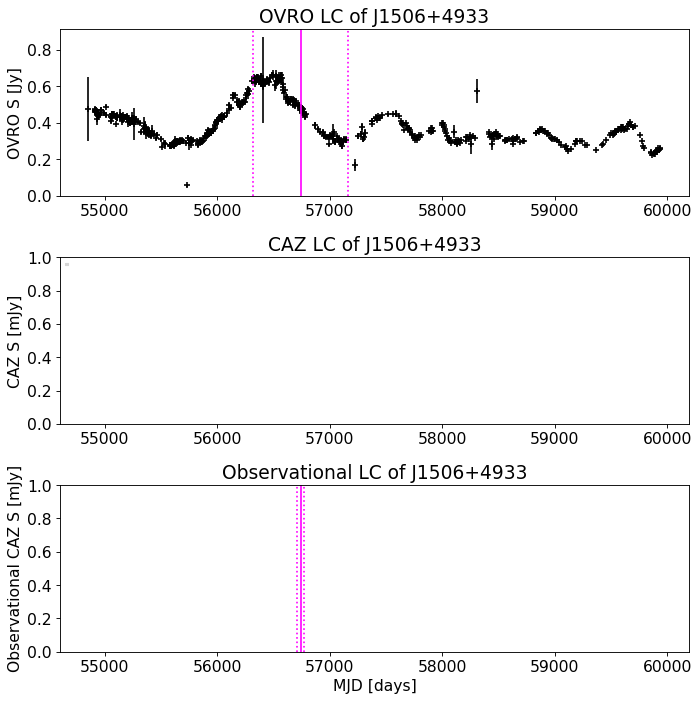}
\includegraphics[width=8cm, keepaspectratio]{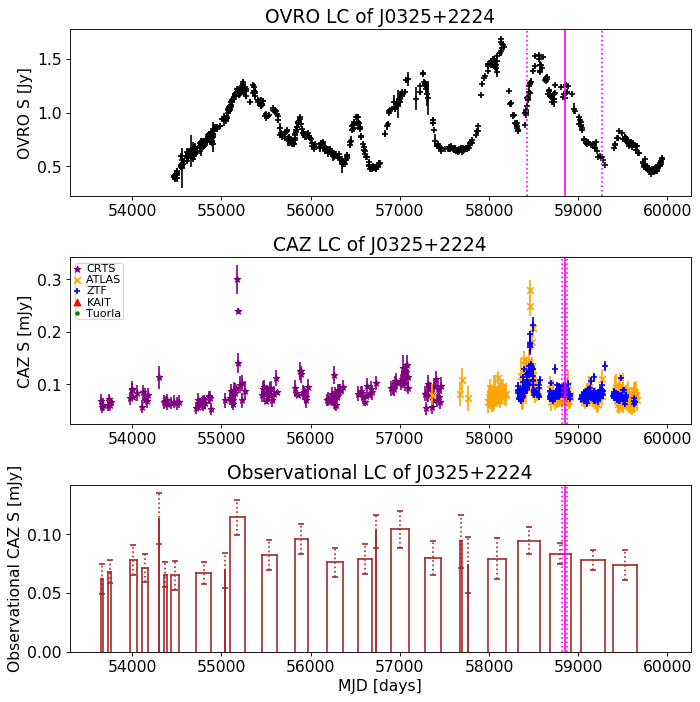}
\\
\raggedright
\textbf{Fig. B.1} continued.
\label{fig:all_flr_LCs4}
\end{figure*}

\begin{figure*}
\centering
\includegraphics[width=8cm, keepaspectratio]{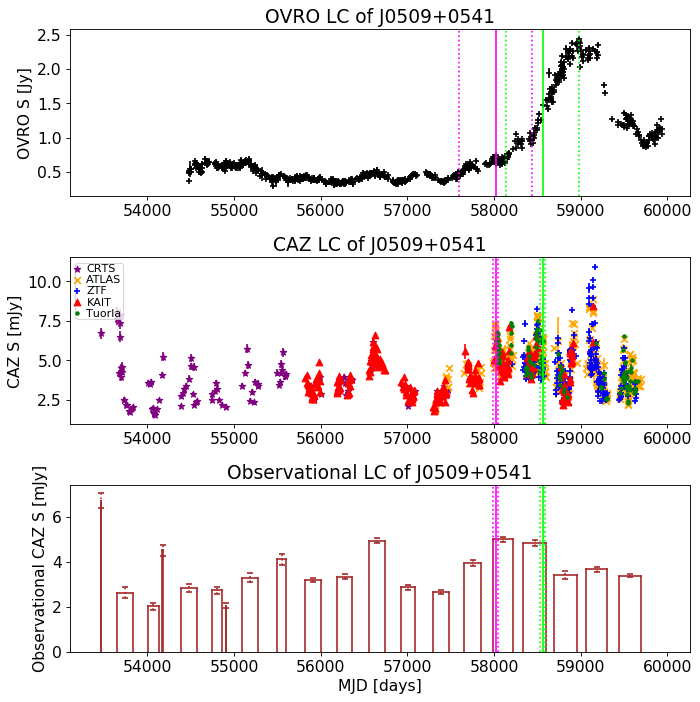}
\includegraphics[width=8cm, keepaspectratio]{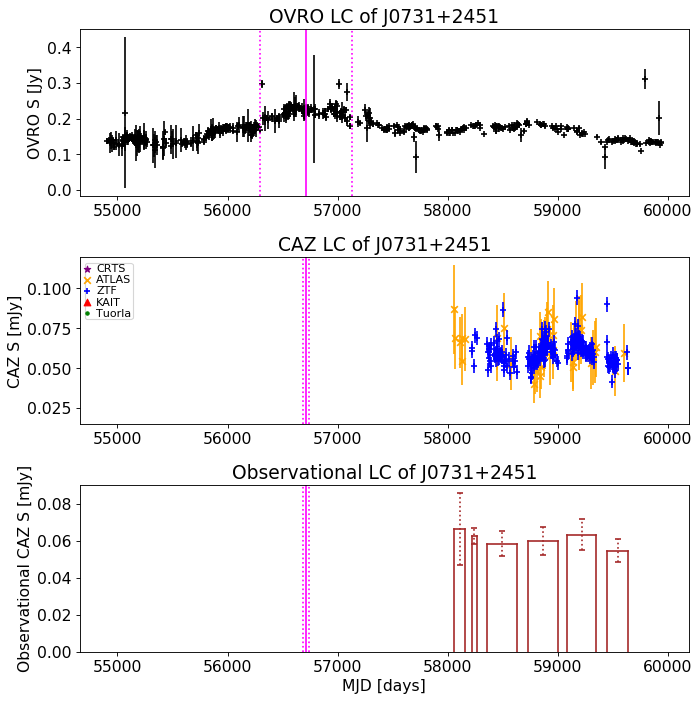}
\includegraphics[width=8cm, keepaspectratio]{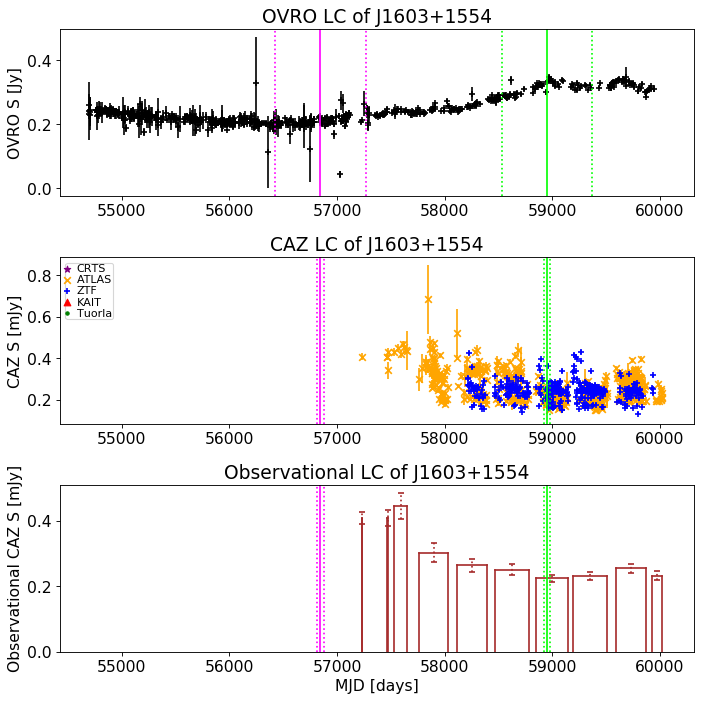}
\includegraphics[width=8cm, keepaspectratio]{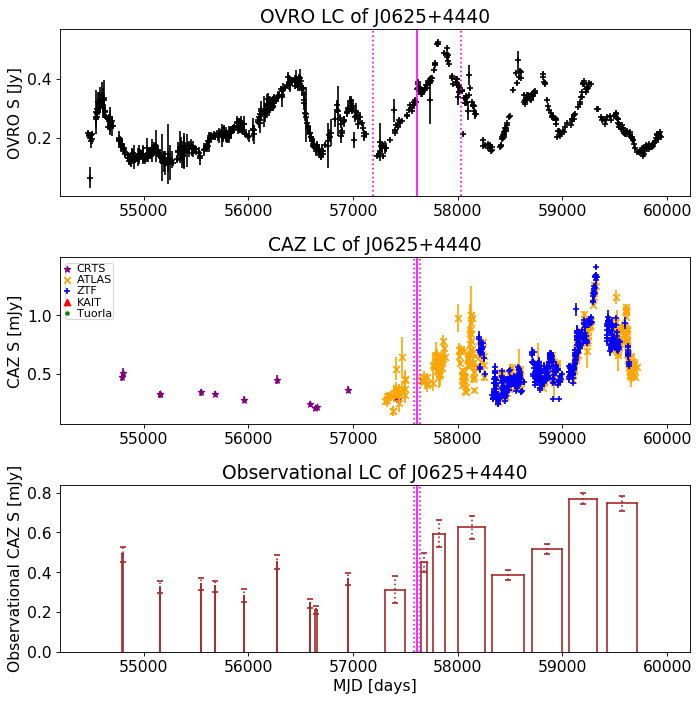}
\includegraphics[width=8cm, keepaspectratio]{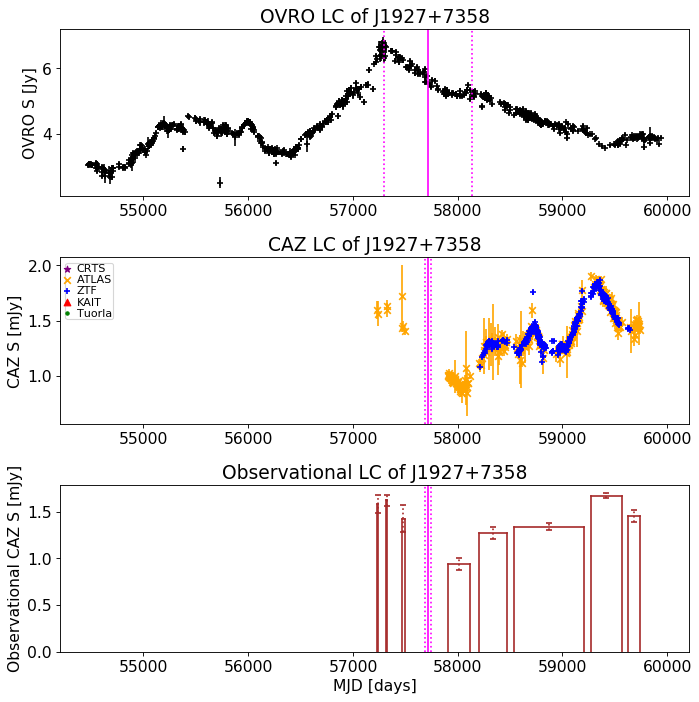}
\includegraphics[width=8cm, keepaspectratio]{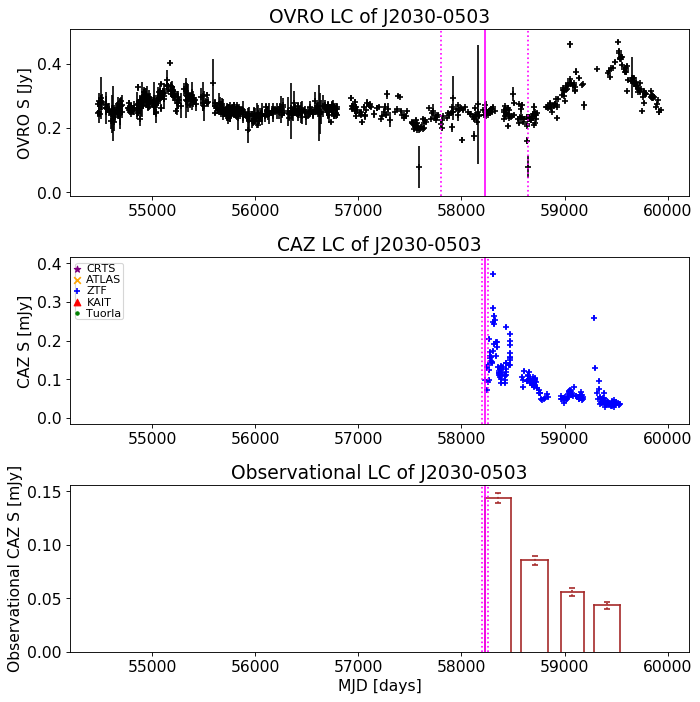}
\\
\raggedright
\textbf{Fig. B.1} continued.
\label{fig:all_flr_LCs5}
\end{figure*}

\begin{figure*}
\centering
\includegraphics[width=8cm, keepaspectratio]{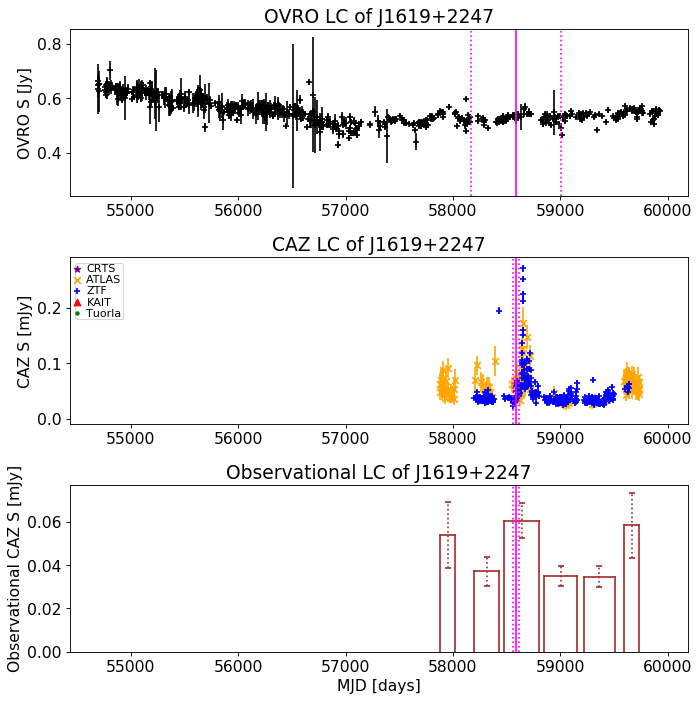}
\includegraphics[width=8cm, keepaspectratio]{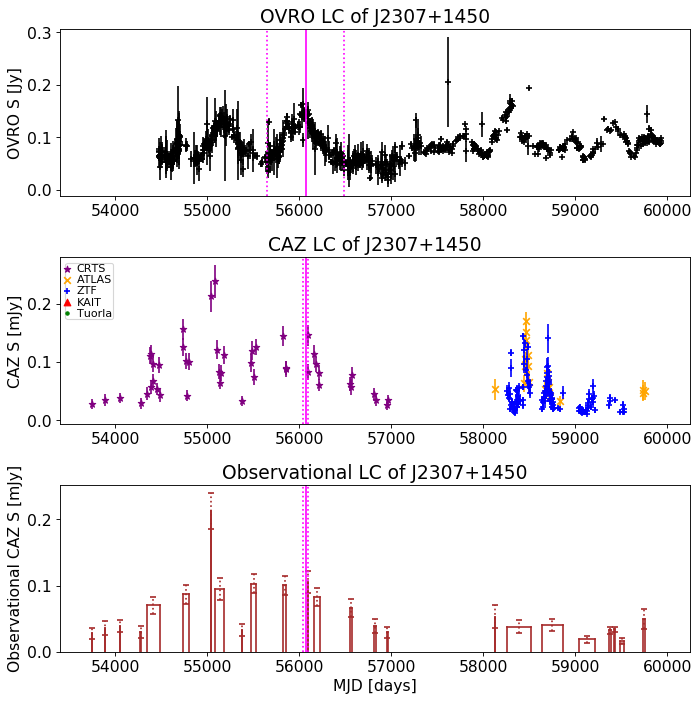}
\includegraphics[width=8cm, keepaspectratio]{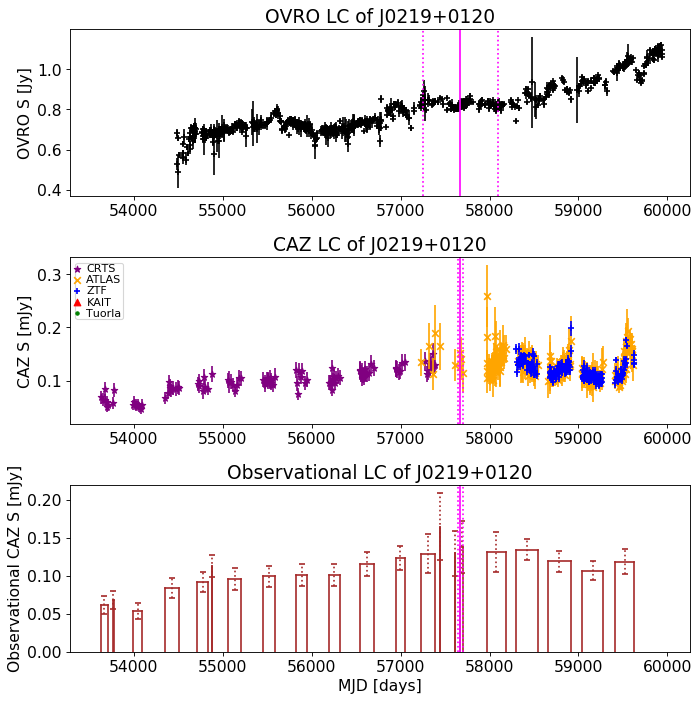}
\includegraphics[width=8cm, keepaspectratio]{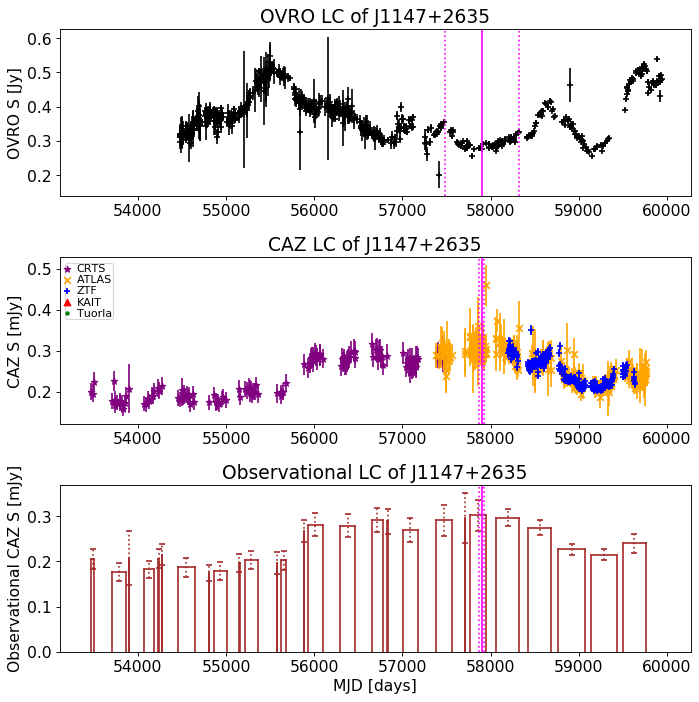}
\includegraphics[width=8cm, keepaspectratio]{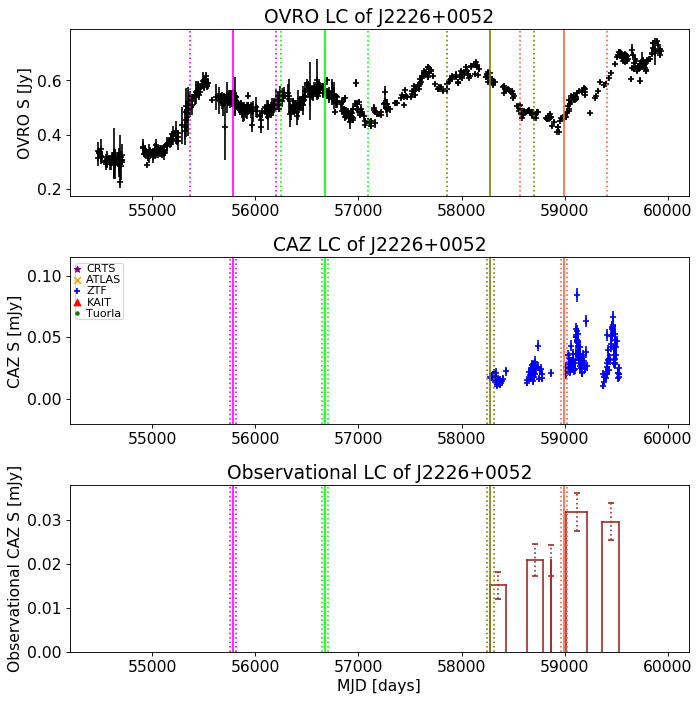}
\includegraphics[width=8cm, keepaspectratio]{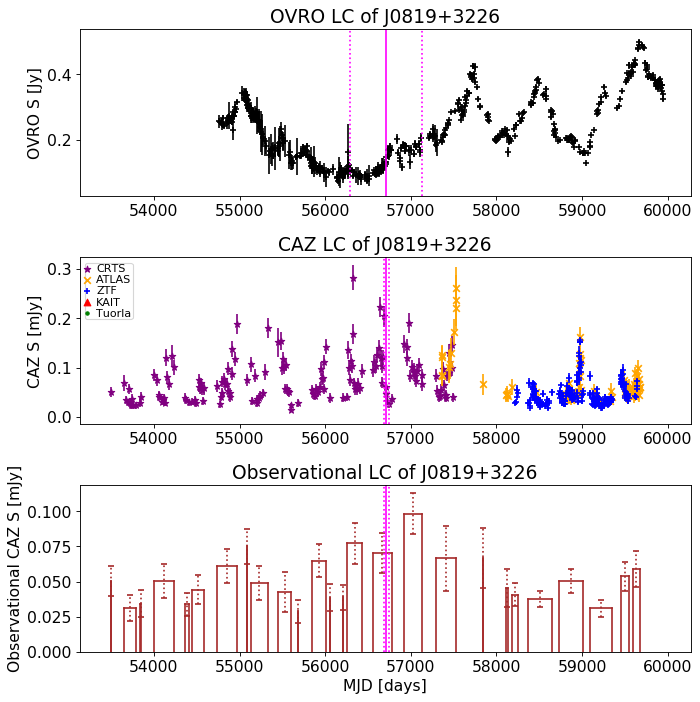}
\\
\raggedright
\textbf{Fig. B.1} continued.
\label{fig:all_flr_LCs6}
\end{figure*}

\begin{figure*}
\centering
\includegraphics[width=8cm, keepaspectratio]{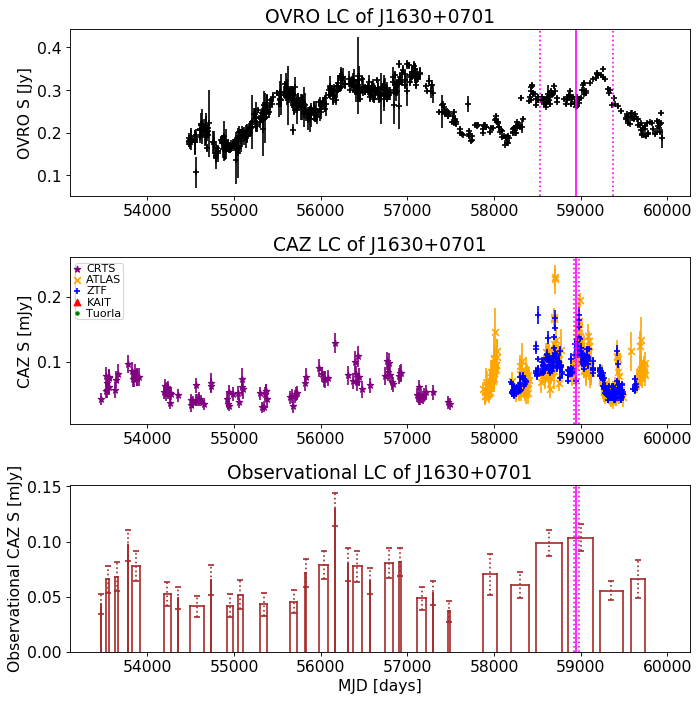}
\includegraphics[width=8cm, keepaspectratio]{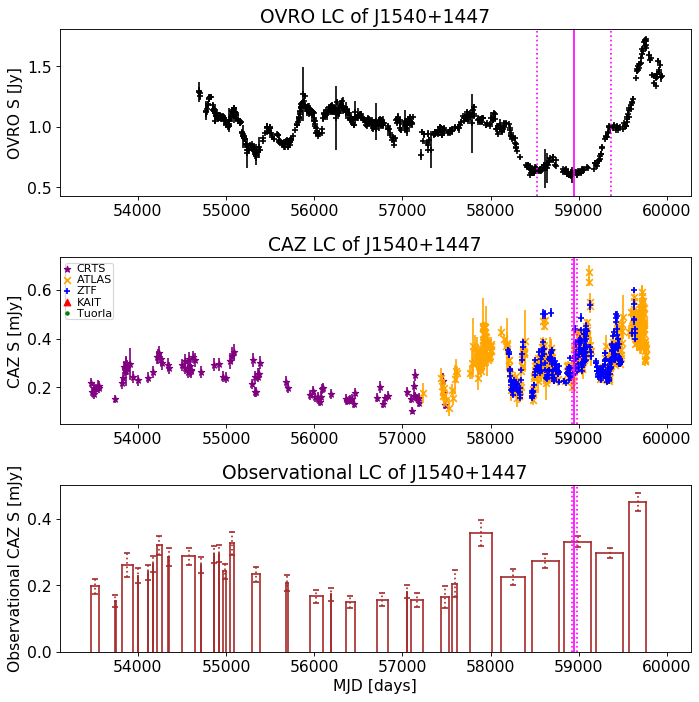}
\includegraphics[width=8cm, keepaspectratio]{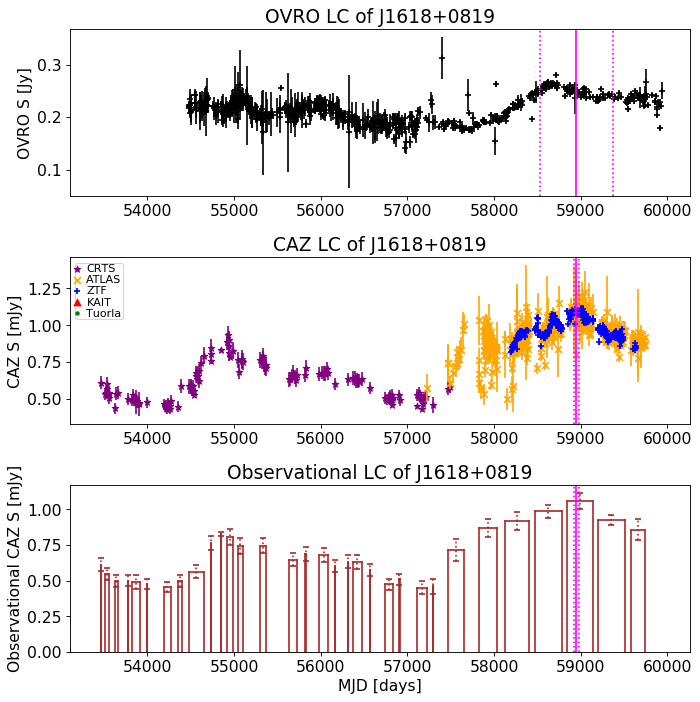}
\\
\raggedright
\textbf{Fig. B.1} continued.
\label{fig:all_flr_LCs7}
\end{figure*}

\end{appendix}

\end{document}